%% file: fies_pub.tex
\documentclass[11pt]{article}

\usepackage{geometry}
\usepackage{fullpage}
\usepackage{graphicx}
\usepackage{pdflscape}
\usepackage{multirow}
\usepackage{ragged2e}

\usepackage{dcolumn}
\usepackage{booktabs,calc}
\usepackage{longtable}
\usepackage{array}
\usepackage{paralist}
\usepackage{verbatim}
\usepackage{subfig}
\usepackage{setspace}
\usepackage{amsmath}
\usepackage{amssymb}
\usepackage{amsthm}
\usepackage{natbib}
\usepackage[linktocpage=true, colorlinks=true, linkcolor=blue, citecolor=black]{hyperref}
\usepackage{enumerate}
\usepackage{authblk}
\usepackage{placeins}
\usepackage{multicol}
\usepackage{array}
\usepackage[usenames, dvipsnames]{xcolor}
\usepackage{url}
\usepackage{standalone}
\usepackage{geometry}
\usepackage{rotating}
\usepackage{pdflscape}
\usepackage{flexisym}
\usepackage{breqn}

\usepackage{verbatim}
\usepackage{amssymb,amsmath,amsthm}
\usepackage[mathscr]{eucal}
\usepackage{makeidx}
\numberwithin{equation}{section}
\usepackage{graphicx}
\usepackage{psfrag}
\usepackage{afterpage}
\usepackage{pdfpages}

\usepackage{hyperref}
\usepackage{ifpdf}
\usepackage{ragged2e}
\usepackage{tikz}

\newcolumntype{J}[1]{>{\justifying\arraybackslash}p{#1}}
\newcolumntype{P}[1]{>{\centering\arraybackslash}m{#1}}
\newcommand{\sym}[1]{\rlap{#1}}

\ifpdf
\fi

    \defcitealias{ETH1}{CSA, 2019}
    \defcitealias{ETH2}{WB, 2022}
    \defcitealias{MWI1}{NSO, 2020}
    \defcitealias{MWI2}{NSO, 2022}
    \defcitealias{NGA1}{NBS, 2019}
    \defcitealias{NGA2}{NBS, 2022}
    \defcitealias{UGA1}{UBOS, 2020}
    \defcitealias{UGA2}{UBOS, 2022}
    
\begin{document}

\title{Coping or Hoping? Livelihood Diversification and Food Insecurity in the COVID-19 Pandemic\thanks{Corresponding author email:  \href{mailto:jdmichler@arizona.edu}{jdmichler@arizona.edu}. Authors are listed alphabetically. A pre-analysis plan for this study was filed prior to completion of post-outbreak data collection at \href{https://osf.io/nu593}{https://osf.io/nu593}. Funding for data collection and analysis comes from the World Bank Multi-Donor Trust Fund for Integrated Household and Agricultural Surveys in Low and Middle-Income Countries (TF072496). The funders had no role in study design, data collection and analysis, decision to publish, or preparation of the manuscript. We acknowledge the research assistance provided by Lorin Rudin-Rush and Joshua Brubaker. We appreciate comments and feedback from participants at the 2021 NIFA Agriculture Policy Conference, the $31^{st}$ International Conference of Agricultural Economists, the 2021 World Bank Development Data Group Learning Series, the Learning from Longitudinal Studies in LMICS: Before, During, and After COVID-19 workshop in 2021, the IFAD Conference 2022, and the Agricultural and Applied Economics Association Annual Meeting in Anaheim. We thank the individuals involved in the design, implementation and dissemination of high-frequency phone surveys on COVID-19, specifically the World Bank LSMS team, and the phone survey managers and interviewers at the Malawi National Statistical Office, the Nigeria Bureau of Statistics, the Uganda Bureau of Statistics and Laterite Ethiopia.}}

	\author[1]{Ann M. Furbush}
	\author[2]{Anna Josephson}
	\author[3]{Talip Kilic}
	\author[2]{Jeffrey D. Michler}
	\affil[1]{\small \emph{Cambridge Econometrics}}
	\affil[2]{\small \emph{Department of Agricultural and Resource Economics, University of Arizona}}
	\affil[3]{\small \emph{Development Data Group (DECDG), World Bank}}

\date{September 2024}
\maketitle

\thispagestyle{empty}

\begin{center}\begin{abstract}
		\noindent We examine the impact of livelihood diversification on food insecurity amid the COVID-19 pandemic. Our analysis uses household panel data from Ethiopia, Malawi, and Nigeria in which the first round was collected immediately prior to the pandemic and extends through multiple rounds of monthly data collection during the pandemic. Using this pre- and post-outbreak data, and guided by a pre-analysis plan, we estimate the causal effect of livelihood diversification on food insecurity. Our results do not support the hypothesis that livelihood diversification boosts household resilience. Though income diversification may serve as an effective coping mechanism for small-scale shocks, we find that for a disaster on the scale of the pandemic this strategy is not effective. Policymakers looking to prepare for the increased occurrence of large-scale disasters will need to grapple with the fact that coping strategies that gave people hope in the past may fail them as they try to cope with the future.
	\end{abstract}\end{center}

	{\small \noindent\emph{JEL Classification}: F6, I38, O2, Q18 \\
	\emph{Keywords}: COVID-19 pandemic, coping strategies, food security, pre-analysis plan, Sub-Saharan Africa}
 
\newpage
\onehalfspacing


\section{Introduction}

The COVID-19 pandemic exacerbated many of the hardships faced by impoverished households. Due to their limited resources, households across Sub-Saharan Africa are particularly vulnerable to shocks, such as natural disasters and pandemics. While households cannot control their risk of exposure, they can employ \textit{ex ante} and \textit{ex post} coping strategies to mitigate the impact of realized risks and enhance their own resilience. For households in Sub-Saharan Africa, among whom formal insurance is uncommon, these coping strategies often involve reallocation of resources, monies, and labor within the family or household \citep{Ellis98}. Households can diversify their incomes to bolster resilience to future shocks and uncertainty - an \emph{ex ante} coping behavior \citep{welderufaelAnalysisHouseholdsVulnerability2014, arslanDiversificationClimateVariability2018, KhanMorrissey23} - or decide to diversify only after a shock exposes their vulnerability - an \emph{ex post} coping behavior \citep{asfawHeterogeneousImpactLivelihood2019, mulwaFarmDiversificationAdaptation2020, cely-santosFightingChangeInteractive2021}.

We study the impact of livelihood diversification on mitigating food insecurity, a proxy for welfare, during the COVID-19 pandemic. To do this, we use household panel data collected by the World Bank, which combines face-to-face survey data collected in Ethiopia, Malawi, and Nigeria prior to the COVID-19 pandemic with monthly post-outbreak phone surveys. This allows us to establish causal relationships to understand how households coped with the impacts of the pandemic through livelihood diversification. Prior to the public release of the phone survey data, we pre-specified our analysis and registered the pre-analysis plan with the Open Science Foundation (OSF) \citep{PAP}. We report on the results of two research questions. First, how has household income composition and livelihood diversification changed since the onset of the pandemic? And second, how does household income composition and livelihood diversification impact household food insecurity amid the pandemic?\footnote{In our pre-analysis plan, we specify a third research question: how do changes in income composition and livelihood diversification and subsequent effects on food insecurity vary across different population subgroups? We discuss findings related to this research question in the Online Appendix.}

With respect to our first question, we interrogate the data and produce stylized facts regarding how livelihood diversification changes during the pandemic relative to pre-pandemic diversity. We do not observe a substantial nor a systematic change in household income composition nor livelihood diversification since the start of the pandemic. We do observe small differences in Ethiopia and Malawi, where households become more specialized during the pandemic than before. These changes are driven by a decline in the percent of households receiving remittances, government assistance, and wage income following the onset of COVID-19. This effect is relatively modest and we cannot disentangle if this was a voluntary coping strategy or an involuntary separation from diverse activities (e.g., the loss of a wage job rather than leaving a wage job). In Nigeria, diversification increased slightly after the start of the pandemic, mainly due to increased participation in farming and greater government assistance. From these modest trends we conclude that households made limited use of livelihood diversification as an \emph{ex post} coping strategy. This may be due to the unique nature of government response to the pandemic, which placed constraints on how households could respond, though these restrictions also did not lead to an across the board increase in the specialization of livelihood activities. 

To answer our second question, we use a dynamic panel model and an ANCOVA estimation to assess changes in household food insecurity. A rich body of literature evaluates livelihood diversification as a both an \emph{ex ante} and \emph{ex post} coping strategy to improve recovery from and resilience to shocks, particularly those related to climate and civil unrest. These studies generally coalesce around the conclusion that income diversification improves household welfare \citep{arslanDiversificationClimateVariability2018, dagungaWhatExtentShould2020, welderufaelAnalysisHouseholdsVulnerability2014}, though there is important heterogeneity based on gender of the household head and whether the household is urban or rural \citep{KhanMorrissey23}. We do not find evidence for this positive relationship in our analysis. Across multiple econometric specifications, combining income sources into various indices, and conducting sub-group analysis by gender and location, nearly all our results are nulls. We then discuss possible explanations for our null results, presenting robustness checks were possible. We conclude that the results are true nulls. The finding that livelihood diversification prior to and during the pandemic has no effect on welfare, as proxied by food insecurity, which runs contrary to much of the previous literature, may be due to the extreme conditions of the COVID-19 pandemic.

This paper contributes to the existing literature on livelihood diversification, as well as to the emergent literature on the impacts and effects of the COVID-19 pandemic. In terms of livelihood diversification, many studies conclude that diversification of income sources reduces poverty and enhances resilience \citep{dagungaWhatExtentShould2020, welderufaelAnalysisHouseholdsVulnerability2014, arslanDiversificationClimateVariability2018, mulwaFarmDiversificationAdaptation2020}. A recent paper examines if diversified firms are more resilient than specialized ones in the face of the COVID-induced market shock \citep{StevensTeal23}. Using food insecurity, which is a common proxy for household welfare in many studies \citep{ebhuomaDefyingOddsClimate2017, guptaDarkBrightSpots2021, harttgenAnalyzingNutritionalImpacts2016, oskorouchiFloodsFoodSecurity2021, wossenImpactsClimateVariability2018}, we build on this body of existing literature to extend our understanding of the role of livelihood diversification in bolstering household resilience to severe socioeconomic shocks like the COVID-19 pandemic. 

Despite the relative recency of the COVID-19 outbreak, there is already a substantial body of literature summarizing the socioeconomic ramifications of the pandemic on households in low-income countries, including on income \citep{RePEc:fpr:ifprid:1979, stoopCovid19VsEbola2021}, well-being \citep{Bauetal22}, food security \citep{HirvonenEtAl21, kansiimeCOVID19ImplicationsHousehold2021, RudinRushetal22}, and other welfare outcomes \citep{furbushEvolvingSocioeconomicImpacts2021, josephsonSocioeconomicImpactsCOVID192021, mahmudHouseholdResponseExtreme2021, FavaraEtAl22}. A subset of this literature focuses on understanding specific policies or transfer programs associated with the pandemic \citep{Bottanetal21, Gulescietal21, Berkouweretal22, Brooksetal22, DietrichEtAl22, Gutierrezetal22} or informational transfer \citep{Bahetyetal21, Sadishetal21}. However, much of this literature remains descriptive in nature and does not investigate changes due to the pandemic, but rather changes occurring during the pandemic. We extend this conversation by estimating causal relationships about coping with the pandemic through livelihood diversification, using both pre- and post-outbreak data. 

Finally, our paper contributes to the small but growing body of research that uses pre-analysis plans in observational studies. While pre-analysis plans are generally associated with studies relying on experimental data \citep{DufloEtAl20}, the first use of a pre-analysis plan in economics was an observational study of the impacts of minimum wage laws \citep{Neumark01}. The main argument against using pre-analysis plans in observational studies is the difficulty in credibly committing to a plan prior to data availability \citep{Olken15}. But, as \cite{JanzenMichler20} argue, there are numerous study settings where research questions can be clearly formulated ahead of the release of data. Democratic elections \citep{HumphreysEtAl13}, policy changes \citep{Neumark01}, and the timed release of government data \citep{ChangEtAl20} are all examples in which researchers combine pre-analysis plans with observational data. In our case, in the month immediately following the outbreak of COVID-19, the World Bank formulated a plan to collect at least 12 rounds of monthly panel data from households that had been surveyed in the year prior to the pandemic. This commitment to future data collection, following a standardized survey instrument, allowed us to formulate hypotheses, develop an empirical approach, and register our plan prior to the collection and public release of all rounds of data \citep{PAP}. In a research setting in which there are numerous ways one could define the variables of interest and model their relationships, a pre-analysis plan lends credibility to our analysis.


\section{Data}

Our analysis focuses on changes to food insecurity after the outbreak of COVID-19 relative to food insecurity status pre-COVID. The spread of the virus impacted household finances indirectly, largely through the closure of businesses and schools and the interruption of supply chains. Governments in the three countries imposed various restrictions to movement, business interactions, and on educational institutions throughout the course of the pandemic. While these restrictions sought to slow the spread of the virus and protect citizens from infection, they disrupted normal activities including household income generation. 


\subsection{COVID-19 Shock}

We describe the circumstances in each country, based on government restrictions that were in place during each data round. In Ethiopia, restrictions were largely implemented at the national-level. Ethiopia closed schools and suspended public gatherings on 16 March 2020. On 8 April 2020, the country declared a state of emergency which included limiting international and domestic travel. However, Ethiopia never went into a complete national lockdown in the sense of closing businesses, restricting movement, or imposing curfews \citep{HirvonenEtAl21}. In Malawi, the President declared a state of disaster on 20 March 2020, which included closing schools and limiting the size of public gatherings. A stay at home order was issued in April. However, this order faced legal challenges, which culminated in the High Court barring the regulation and preventing the stay-at-home order from going into effect, leaving daily economic activity largely intact \citep{josephsonSocioeconomicImpactsCOVID192021}. Nigeria’s response primarily occurred at the state-level. Most Nigerian states closed schools and suspended large gatherings by 24 March 2020 and suspended inter-state travel on 23 April. While non-essential shops as well as restaurants were ordered to close, the government's attempts to impose these closures along with curfews, social distancing, and self-quarantine, were largely ignored, meaning daily economic activity was relatively unchanged \citep{JacobsOkeke22}. The government lifted the closure order less than a month later. Compared to lockdowns in China, Europe, and the United States, the closures of businesses and the restrictions to daily activities in Ethiopia, Malawi, and Nigeria were substantially less strict. This is important since it means that households in these countries were less constrained in pursuing livelihood opportunities than those in regions of the world with government imposed lockdowns.

\begin{figure}[!htbp]
	\begin{minipage}{\linewidth}		
		\caption{COVID-19 Government Restriction Stringency Score Over Time}
		\label{fig:stringency}
		\begin{center}
    	\includegraphics[width=\linewidth,keepaspectratio]{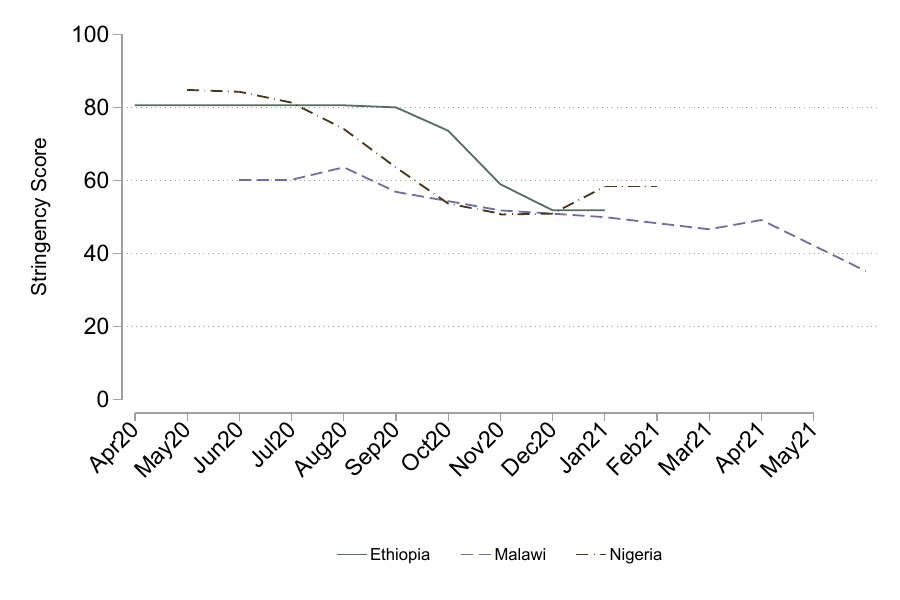}
		\end{center}
		\footnotesize  \textit{Note}: The figure presents government stringency scores for each of the three countries over time. The scores are provided by \textit{Our World in Data} and measure the severity of COVID-19-related government restrictions on a daily basis, with higher scores indicating stricter regulatory regimes \citep{ritchieCoronavirusPandemicCOVID192020}. We average these daily scores to match with our monthly data. In general, government restrictions were most stringent in early 2020 and relaxed in late 2020. In some cases, new restrictions were imposed in early 2021 as cases surged. 
	\end{minipage}	
\end{figure}

To account for the variation in COVID-19-related restrictions over time, we use \textit{Our World in Data}'s COVID-19 Government Stringency Index in some of our empirical specifications \citep{ritchieCoronavirusPandemicCOVID192020}. The index considers nine metrics to calculate daily scores for each country: school closures; workplace closures; cancellation of public events; restrictions on public gatherings; closures of public transport; stay-at-home requirements; public information campaigns; restrictions on internal movements; and international travel controls. The stringency index is calculated as the mean score of the nine metrics, each taking a value between 0 and 100. A higher score indicates a stricter regulatory regime. To match these daily data to each round of our data, we take the average daily score during each survey period. Figure \ref{fig:stringency} displays the average government stringency index in each country over time. 


\subsection{Sample Selection and Surveys} 

To examine the relationship between livelihood diversification and welfare outcomes, we use panel data from high frequency phone surveys (HFPS) in Ethiopia, Malawi, and Nigeria \citepalias{ETH2, MWI2, NGA2}. In each country, interviewers conduct these surveys on a monthly basis with households for a period of at least 12 months following the outbreak of COVID-19.\footnote{The following agencies implement the monthly surveys with support from the World Bank Living Standards Measurement Study (LSMS): Laterite Ethiopia, the Malawi National Statistical Office, and the Nigeria Bureau of Statistics.} The sample for the HFPS is drawn from households that had been interviewed during the most recent (2019) round of the national longitudinal household survey implemented by the respective national statistical office, with assistance from the World Bank \citepalias{ETH1, MWI1, NGA1}. These pre-COVID-19 Living Standards Measurement Study - Integrated Surveys on Agriculture (LSMS-ISA) data are representative at the national, regional, and urban/rural levels and serve as a baseline for our post-COVID-19 analysis.

The HFPS are not nationally representative as participation requires that each household have (1) at least one member who owns a phone, (2) cell network coverage, and (3) access to electricity. These requirements may lead to selection bias in the survey sample. Additionally, the surveys may suffer from non-response bias if targeted households were not willing or able to participate. To address these challenges, we use survey weights provided in the HFPS data which include selection bias corrections and post-stratification adjustments. Several studies using the HFPS data have found that the use of survey weights and post-stratification adjustments substantially reduce the bias, though it does not fully eradicate the bias, and our results should be interpreted with this in mind \citep{AmbelEtAl21, BrubakerEtAl21, GourlayEtAl21}. For a detailed description of the weight calculations used in this study, see \cite{josephsonSocioeconomicImpactsCOVID192021}.

The integration of data from the post-outbreak HFPS and pre-outbreak LSMS-ISA surveys allows us to capture the variation in the effects of the pandemic across a diverse set of Sub-Saharan Africa countries and over time. Importantly, the combined data afford us the opportunity to examine the effects of COVID-19 in relation to a pre-pandemic baseline, allowing us to establish a causal relationship between our variables of interest. The surveys feature cross-country comparable questionnaires on a range of topics including participation in income-generating activities and food insecurity. In total, over 9,000 households are included in this analysis. With baseline LSMS-ISA data in all three countries plus 10 rounds of HFPS data in Ethiopia and 11 in Malawi and Nigeria, our research draws from a total of over 34,000 observations. The average number of households in each round of data is: 2,784 in Ethiopia, 1,611 in Malawi, and 1,943 in Nigeria, though the actual number of households in the baseline and each round differ due to attrition.


\subsection{Livelihood Diversification Indices}

Prior to describing how we construct each of our livelihood diversification indices, we present the disaggregated categories in which households engage in income generation (see Table~\ref{incsum}). The table shows pre-COVID income sources in each country at the most detailed level provided by the classification system used by the World Bank in the LSMS surveys.\footnote{In the LSMS data, income is reported in the local currency. To allow for cross-country comparisons, we convert income values to constant 2019 US dollars.} Income data is collected for all household members and we have summed across individuals to construct a household-level measure. The indices we consider in this paper use aggregated versions of these income sources, though we show the most detailed level here for completeness.

The post-outbreak phone surveys collected less detailed income data than the pre-outbreak in-person surveys. The income module was also not asked in every survey round. In surveys that did included a module on sources of income, those modules asked ``In the last 3 months, which of the following were your household's sources of livelihood?'' Options included: 1) Family farming, livestock or fishing; 2) Non-farm family business, including family business; 3) Wage employment of household members; 4) Remittances from abroad; 5) Assistance from family within the country; 6) Assistance from other non-family individuals; 7) Income from properties, investments or savings; 8) Pension; 9) Assistance from the Government; 10) Assistance from NGOs / charitable organization/religious bodies; 11) Other income source. A household is assigned a one if it reports ``yes'' to the question and zero if it reports ``no.''

Our variable of interest in our analysis consists of a series of indices measuring income diversification. Following methodology from \cite{michlerSpecializeDiversifyAgricultural2017}, we use two measures to evaluate household income diversification: (1) a simple fractional index (FI) and (2) a Herfindahl-Hirschman Index (HHI).\footnote{We also generate four variations of these indices, following our pre-analysis plan, that differ in if they standardize income categories across countries and/or across time. Summaries of these additional indices are presented in Online Appendix~\ref{sec:data_app}.} HHI scores are negatively related to diversification. That is, they are larger for less diversified households and smaller for more diversified households. For consistency, we adjust the fractional index to maintain this negative relationship. Both indices can be interpreted as specialization indices that are inversely related to diversification. Table~\ref{tab:indices_main} summarizes the characteristics of each index.

The simple fractional index is calculated using the count of the income sources each household is engaged in ($m$) at time $t$ given the total number of income-generating opportunities ($n$) in their region ($j$) over the entire time series:

\begin{equation} \label{eq:fraction}
FI_{ijt} = 1-\frac{m_{it}}{n_{j}}.
\end{equation}

\noindent The fraction is subtracted from one so that a higher score is associated with fewer income-generating activities, while a lower score indicates a more diversified income portfolio. The fractional index is calculated for both pre- and post-COVID-19 data. To generate the index, we collapse multiple income categories into seven categories standardized across the three countries: (1) farm, (2) wage, (3) pension, (4) remittances, (5) non-farm enterprises, (6) income from properties, savings, and investment, and (7) other, which includes asset sales, income from NGOs, and other government assistance. This index accounts for geographic area to capture livelihood specialization relative to regional diversification opportunities. We count the total number of income sources households participate in for all geographic areas available in the data (e.g., region, zone, district, postal code, ward). We then determine the smallest geographic area with at least 10 available observations. The count of income sources households are engaged in within that smallest geographic area with sufficient observations then serves as the denominator ($n_j$) in the index calculation for households residing in that area.\footnote{Our final results are robust to changes in this benchmark: if we include a control for the size of the benchmark or standardize the geographic region that is the benchmark for diversification we see no difference in how diversification relates to food insecurity.} 

The HHI considers the portion of a household's income generated from each income source. In calculating the HHI, we include all revenue generated by households but do not net out costs of production. The HHI is calculated as: 

\begin{equation} \label{eq:HHI}
    HHI_{i} = \sum_{m=1}^{M} p_m^{2},
\end{equation}

\noindent where $M$ represents each household's total number of income sources. Each $p_m$ represents the percentage of the household's income generated from income source $m$. A highly specialized household with only one income source would receive the highest possible score of 1 ($1^2$). Similarly, a household with two income sources each accounting for 50 percent of household's total income would receive a score of .5 ($.5^2+.5^2$). As with the simple fractional index, higher scores indicate more income specialization and less diversification. The HHI includes only pre-COVID data and so there are no time sub-scripts. To generate the index, we use 12 income categories standardized across countries, and the respective amount of income each household earns from each source. These categories are: (1) remittances, (2) in-kind assistance from family and friends, (3) investments and savings, (4) income from properties, (5) pension, (6) non-farm enterprises, (7) crop sales and consumption, (8) livestock sales, (9) livestock product sales and consumption, (10) wages, (11) government and NGO assistance, and (12) other.

Before proceeding, it is necessary to comment on the limitations of our measure. First, when it comes to survey-based measures of household finances, asking questions about consumption is generally preferred to asking questions about earned income. Wealthier households tend to under-report earned income \citep{CarlettoEtAl22}. In this study, however, we examine specifically the sources of income, and so a consumption approach would provide no information on the sources from which a household earns its livelihood. Second, diversification as a coping strategy is not about how many different sources of income one has but about spreading the risk by diversifying to income sources with different risk profiles. To that end, one would want to have information on crop choice and the sectoral composition of wages. Unfortunately, the data, particularly the HFPS data, does not capture this information. Third, the way in which we have measured diversification cannot distinguish between voluntary and involuntary changes in diversification. For example, a person may be separated from a job, which would appear to be a decrease in diversification, though an involuntary one, not a deliberate coping strategy. Ideally, we could differentiate between involuntary and voluntary diversification actions, but we can only observe the level of diversification and the subsequently related associated food insecurity. Finally, as job searches, crop production, and starting a new business typically extend over several months, one would want to have data over a long enough time frame to adequately allow for inter-household adjustments to livelihood sources. While our data span a time period of more than two years, we acknowledge that this time frame is shorter than most other studies of livelihood diversification, some of which span decades \citep{michlerSpecializeDiversifyAgricultural2017, KhanMorrissey23}.


\subsection{Food Insecurity}

We examine food insecurity as our primary outcome variable to measure household well-being. We use the Food Insecurity Experience Scale (FIES), which is an experience-based metric which can be used to compare prevalence rates of food insecurity across national and sub-national populations. Following the FIES standard survey model \citep{SmithEtAl2017}, respondents to the pre- and post-outbreak surveys answer eight questions aimed to capture whether the respondent or other adult households members: 

\begin{enumerate}
    \item were worried they would not have enough to eat,
    \item were unable to eat healthy and nutritious food,
    \item ate only a few kinds of food,
    \item had to skip a meal,
    \item ate less than they thought they should,
    \item ran out of food,
    \item were hungry but did not eat, or
    \item went without eating for a whole day.
\end{enumerate}

\noindent Following standard practice \citep{bloemCOVID19WorkingPaper2021}, we count the number of affirmative answers to these eight questions to categorize households into mild, moderate, and severe food insecurity. Households which answered affirmatively to between one and three FIES questions are classified as experiencing mild food insecurity. Households which answered yes to between four and seven questions are classified as experiencing moderate food insecure. Households are classified as severely food insecure if they responded affirmatively to all eight questions.

FIES scores using these integer values may be limited by several factors. First, some post-outbreak rounds do not include food insecurity modules, so there are gaps in the data, just as there are with income sources. Second, there are inconsistencies in the reference period for food insecurity questions in the pre-outbreak data.\footnote{Online Appendix~\ref{sec:data_app} includes a the exact wording of each question (including reference periods) in each country both before and after the pandemic began.} In Ethiopia and Malawi, the reference period in the pre-outbreak data is the last seven days while in the post-outbreak period it is 30 days. In Nigeria, the reference period is consistent (30 days) both pre- and post-outbreak. Finally, in Malawi, only six of the eight questions were included in the pre-outbreak survey.

To ensure our measures of food insecurity are as similar as possible over time, we create a standardized FIES score developed by \cite{ADJOGNON2021102050} and implemented in \cite{RudinRushetal22} that is in addition to our mild, moderate, and severe indicators. The standardized measure counts the number of affirmative answers to FIES questions in the pre-outbreak data by country and uses survey weights to standardize the variable such that its mean is zero and standard deviation is one.\footnote{The standardization process creates a $z$-score by subtracting off the mean and dividing by the standard deviation to get $FIES \sim \mathcal{N}(0,1)$. In calculating the mean and standard deviation, we use weights so that the mean is the weighted mean and the standard deviation is the weighted standard deviation. This preserves the representativeness in the standardized variable.} Following a similar process, the post-outbreak data are standardized by country across all data rounds. As such, the standardization process facilitates comparison between pre- and post-outbreak data and across country by ensuring our measure of food insecurity is as similar as possible over time and across countries. This allows for comparisons of deviations from the pre-pandemic mean and the mean of the variables after the onset of the pandemic within each country. Additionally, standardization allows us to interpret estimated coefficients in terms of standard deviations instead of a unitless score.

\begin{figure}[htbp]
    \begin{minipage}{\linewidth}
         \caption{Food Insecurity Measures Over Time}
        \label{fig:fies}
        \begin{center}
            \includegraphics[width=\linewidth,keepaspectratio]{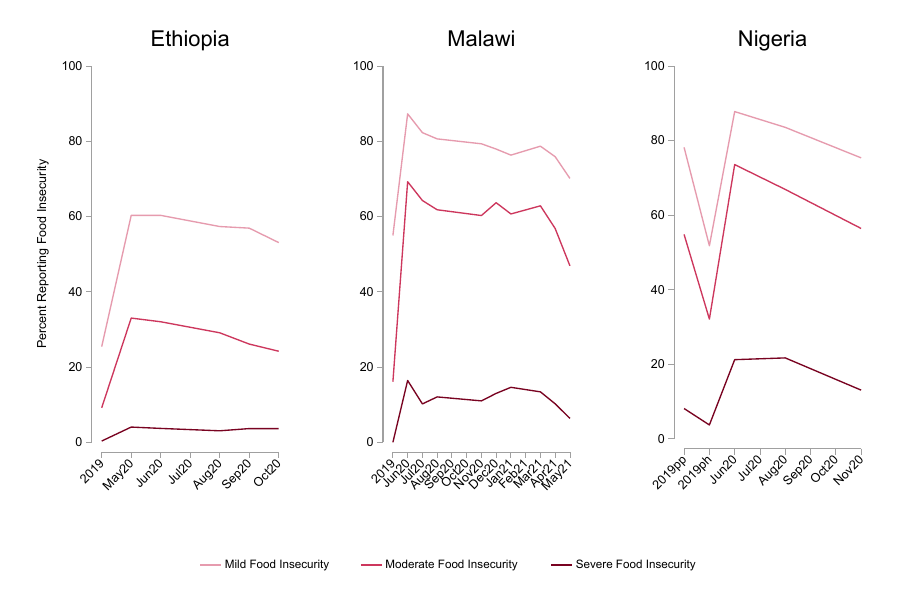}
         \end{center}
        \footnotesize \textit{Note}: The figure shows the percentage of households experiencing mild, moderate, and severe food insecurity in each round of available data. In Nigeria, the pre-outbreak surveys ask respondents food security questions in both a post-planting (labeled ``pp'') and a post-harvest (labeled ``ph'') survey. 
    \end{minipage}
\end{figure}

As seen in Figure \ref{fig:fies}, food insecurity in all three countries increased substantially between 2019 and the summer of 2020, following the onset of the pandemic. Recovery in food security throughout the subsequent year was slow in all three countries, with about 80 percent of households in Malawi and Nigeria and about 60 percent in Ethiopia experiencing mild food insecurity in almost every month following the outbreak. Prior to the onset of the pandemic, mild food insecurity affected less than 30 percent of households in Ethiopia and about 60 percent in Malawi and Nigeria. Similarly, moderate food insecurity spiked after the COVID-19 outbreak and slowly recovered in subsequent months. Severe food insecurity increased in Malawi and Nigeria in June 2020 and rose slightly in Ethiopia after the initial outbreak period. In all three countries, food security has not returned to pre-pandemic levels.


\section{Method} \label{sec:method}

We use two econometric approaches to investigate the causal effect of livelihood diversification on mitigating the impact of COVID-19 on food insecurity. The first approach is a dynamic panel data estimator in which food insecurity for a particular round is explained by the diversity index from the previous round. This approach is designed to capture the potential \emph{ex post} effects of changing livelihood diversification strategies in response to the pandemic. The second approach is an ANCOVA estimator, in which we regress food insecurity for a particular round on the pre-COVID-19 diversity index. This approach is designed to capture the potential \emph{ex ante} effects of livelihood diversification in anticipation of a shock. For each specification, we run regressions for each country separately. Survey weights are included in all specifications. 

Following \cite{GilesMurtazashvili13}, who investigate poverty outcomes in linear and non-linear settings, our dynamic panel data model with lagged variables takes the following form:

\begin{equation}\label{eq:dyn}
    y_{it}= \alpha + \beta_1 y_{it-1} + \beta_2 (y_{it-1}*div_{it-1}) + \beta_3 div_{it-1} + \delta_t + r_j*t_t + u_i + \epsilon_{it},
\end{equation}

\noindent where $y_{it}$ is food insecurity for household $i$ at time $t$. $y_{it-1}$ is the lagged value of food insecurity and $div_{it-1}$ is the lagged value of the diversity index. We lag these values to account for the time it takes for livelihood diversification to actually affect welfare. Diversification does not have an instant impact. Rather, households may use diversification, \emph{ex ante}, to cope with shocks and thus improve food security in the future. Including lagged food insecurity $y_{it-1}$ in our specifications ensures that the variation we observe in our dependent variable is due to livelihood diversification rather than household-level characteristics or differences. In this specification, $\beta_2$ is our variable of interest, measuring how lagged income diversification impacts a household's food insecurity, dependent on that household's food insecurity status in the prior round of data.\footnote{We pre-specified using past income level and focused on diversification as the coping strategy but our results are robust to alternative proxies, such changes in income level, and use of other coping strategies and alternative estimators. See the Online Appendix~\ref{app:main} for additional details and results.}

In estimation, we account for regional and time differences in COVID-19 policies and mitigation strategies. We include time (i.e., round) indicators $(\delta_t)$ to capture variation in COVID-19 cases and COVID-19-related policies occurring nation-wide. This also captures other large-scale temporal events such as elections in Malawi and civil unrest in Ethiopia. Regional indicators $(r_j)$ are interacted with a time trend $t_t$ to control for regional differences in COVID-19 mitigation strategies over the evolution of the pandemic as well as other regional shocks such as drought or conflict. Lastly, $u_i$ is a household fixed effect to control for time-invariant, unobservable household heterogeneity, and $\epsilon_{it}$ is an idiosyncratic error term. Robust standard errors are clustered by household to correct for within-household correlation over time.

The dynamic panel data specification captures causal impacts if the livelihood diversification index is not correlated with the error term. We address issues of endogeneity which might arise from (1) simultaneity or reverse causality between the independent and dependent variables, (2) omitted variable bias, or (3) non-classical measurement error. With regard to the first issue, it is possible that households with better welfare outcomes have more resources, enabling them to diversify their livelihoods. In this case, a simultaneity problem arises as the dependent and independent variables are co-determined. To account for simultaneity, we use lagged independent variables to ensure temporal precedence of livelihood diversification relative to observed household food insecurity, thus isolating the relationship between past diversification and subsequent outcomes.\footnote{The use of lagged variables has the potential to bias coefficients downward, which \cite{ArellanoBond91} and others address through instrumental variables. In Online Appendix~\ref{app:main} we implement the Arellano-Blover/Blundell-Bond \citep{ArellanoBover95, BlundellBond98} GMM estimator and show that the size of the downward bias is small and does not affect the interpretation of our results.} Our panel data, with a pre-COVID baseline helps us avoid omitted variable bias. By including household fixed effects, we account for observable and unobservable time-invariant household characteristics that might influence food insecurity. Including time dummies and region-time trends control for unobserved time varying heterogeneity, albeit not at the household-level. As a result, we greatly reduce the probability of omitting crucial variables. Finally, we address potential non-classical measurement error by pre-specifying multiple measures of livelihood diversification that categorize and aggregate income in different ways. All robustness checks are presented in Online Appendix~\ref{app:main}. While we cannot account for all possible sources of endogeneity, such as contemporaneous, time-varying, idiosyncratic shocks to the household, our dynamic model, series of controls, and multiple livelihood diversification measures reduce the likelihood of correlation between dependent and independent variables and enable us to credibly claim to identify causal relationships.

To enrich the dynamic panel data model, we add interaction effects to account for the socioeconomic impacts associated with COVID-19 government restrictions:

\begin{dmath}\label{eq:dynint}
    y_{it}=\alpha+\beta_1y_{it-1} + \beta_2div_{it-1}+\beta_3str_t + \beta_4(y_{it-1}*div_{it-1}) + 
    \beta_5(y_{it-1}*str_t) + 
    \beta_6(div_{it-1}*str_t) + \beta_7(y_{it-1}*div_{it-1}*str_t) +  \delta_t+r_j*t_t+u_i+\epsilon_{it}.
\end{dmath}

\noindent Here $str_t$ is the government stringency score at time $t$. The triple interaction term ($\beta_7$) indicates the combined impact of lagged food insecurity, lagged income diversity, and contemporaneous government stringency. As with Equation~\eqref{eq:dyn}, the specification of Equation~\eqref{eq:dynint} includes household fixed effects, time dummies, region-time trends, and clustered standard errors by household.

In addition to the dynamic panel data models, we use an ANCOVA estimator to generate difference-in-difference-type estimates \citep{McKenzie12}.\footnote{We also pre-specified the use of simple difference-in-difference (DID) models in which we include an indicator for the start of the lockdown. We prefer the ANCOVA specifications to simple DID because coefficients are more precisely estimated. However, we present DID results in Online Appendix~\ref{app:main}.} Here we explicitly control for pre-pandemic welfare:

\begin{equation}\label{eq:anc}
 y_{it} = \alpha + \beta_1 y_{it=0} + \beta_2 (y_{it=0}*div_{it=0}) + \beta_3 div_{it=0} + \delta_t + r_j*t_t + \epsilon_{it}.
\end{equation}

\noindent In this equation, $div_{it=0}$ and $y_{it=0}$ are the diversity index and food insecurity in the pre-COVID-19 data. All other terms are as previously defined. Including the pre-COVID outcome variable $y_{it=0}$ more precisely attributes the variation in food insecurity to our variable of interest. In this model, we evaluate the impact of \emph{ex ante} income diversification on post-shock food insecurity. In this context, $\beta_2$ is the variable of interest, which is the relationship between pre-COVID-19 income diversification and household food insecurity during the pandemic, dependent on that household’s food insecurity status before the pandemic. As with the other models, we include time dummies, region-time trends, and cluster standard errors at the household-level.

With respect to identification, by observing the impacts of pre-COVID diversification on post-outbreak food insecurity, we avoid simultaneity issues and potential reverse causality. As we compare within-household changes relative to the baseline, the ANCOVA estimator controls for unobserved time-invariant heterogeneity between households. Additionally, our set of controls capture geographic and temporal shocks that evolve over the pandemic. That said, it is still possible that baseline livelihood diversification affects the probability of a household suffering from a time-varying, idiosyncratic shock at some point during the pandemic that impacts household welfare.


\section{Results}


\subsection{Income Composition Over Time}

We describe a set of stylized facts using graphical analysis and non-parametric regressions in order to evaluate changes in income composition over time. Recall our first research question is: how has household income sources for households and household diversification changed since the onset of the pandemic?

\begin{landscape}
\begin{figure} [htbp]
    \begin{minipage}{\linewidth}
        \caption{Fractional Index: Income Sources Over Time}
        \label{fig:ind1_sources_time}
        \begin{center}
            \includegraphics[width=.9\linewidth,keepaspectratio]{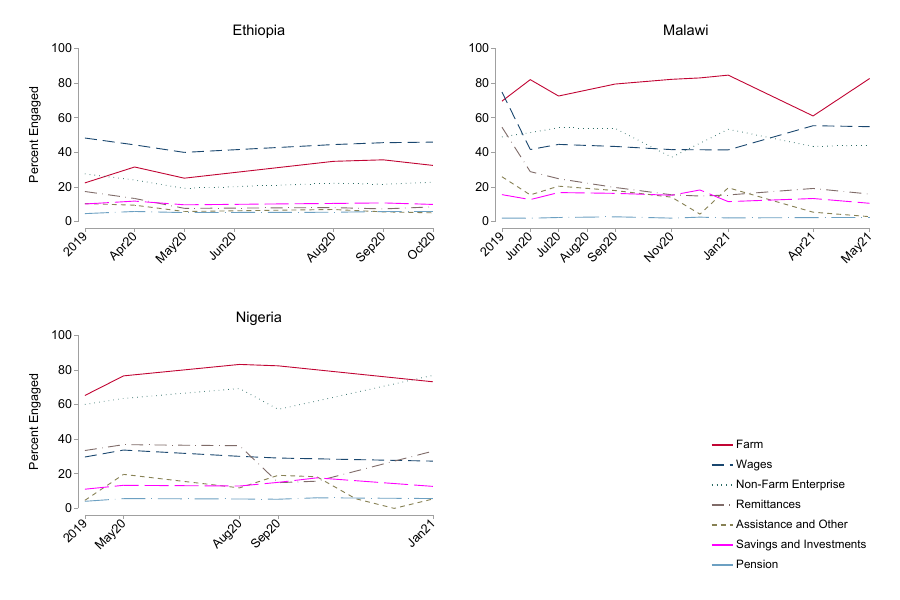}
         \end{center}
        \footnotesize  \textit{Note}: The figure shows the percentage of households earning income in each of the seven categories used to generate the standardized pre- and post-outbreak fractional index by country and by round.
    \end{minipage}
\end{figure}
\end{landscape}

Figure~\ref{fig:ind1_sources_time} reports on changes in the mean number of households earning income from each of the seven income categories included in our fractional index. A first stylized fact that emerges from these non-parametric regressions is that households in all three countries are fairly specialized. While the data are weighted so as to be nationally representative, a majority of households are engage in farming. In Ethiopia, farming far dominates other sources of income. Wages and non-farm enterprises are the next most common sources of income in both countries, though less than 40 percent of household earn income from these sources. In Malawi, over 80 percent of the population earn income from farming, though unlike in Ethiopia, a majority of household report earning income from wages and remittances. In Nigeria, nearly as many households (60 percent) report earning income from farming as from non-farm enterprises.

A second stylized fact is that in most countries there was a small but significant increase during the pandemic in the number of households engaged in farming. In Malawi and Nigeria the number of households reporting income from farming increased after the start of the lockdowns and then slowly decreased over the length of the pandemic. Some of this change is likely due to seasonality, as June and July are harvest season for maize and beans in these countries. However the agro-climatic variation in each country, their location in the tropics, and the diversified nature of smallholder agriculture means that seasonality is unlikely to account for all the temporal change in farm income. A second likely reason for the increase in farm income is household adaptation to business closures and travel restrictions. In fact, we see declines in wage labor and/or remittances that are contemporaneous with the increase in farm income in each of the countries. Ethiopia is the outlier, with farm income decreasing after the pandemic. Again seasonality offers a partial explanation: harvest for crops grown during the long-rains (\emph{mehir}) is November and December, though crops grown during the short-rains (\emph{belg}) is in May and June, when we see participation in farm income declining. The decline in farm income as well as the contemporaneous decline in wage and non-farm enterprises suggests households in Ethiopia became more specialized as a result of seasonal labor demands and government restrictions related to COVID-19.

A third stylized fact is that most households in most countries lost sources of income during the pandemic, particularly in its first months. In Ethiopia, we see a decline in the share of households reporting income from farming, wages, non-farm enterprises, remittances, and assistance. In Malawi, we see declines in wages, remittances, and assistance. That Ethiopia sees a decline in participation from almost every source again suggests a tend towards specialization in response to the pandemic. In Malawi, while households stopped receiving income from various sources, the increase in participation in farming means that the direction of a household's response is indeterminate. Relative to these two countries, Nigeria is an outlier. While remittance income declined, more households report participating in farming, wage labor, and receiving assistance after the start of the pandemic. There was no apparent immediate effect of the pandemic on non-farm enterprise. All of this suggests that household in Nigeria responded to the pandemic by becoming more diversified.

\begin{figure} [htbp]
    \begin{minipage}{\linewidth}
        \caption{Fractional Index Over Time}
        \label{fig:stp_pp_index_time}
        \begin{center}
            \includegraphics[width=\linewidth,keepaspectratio]{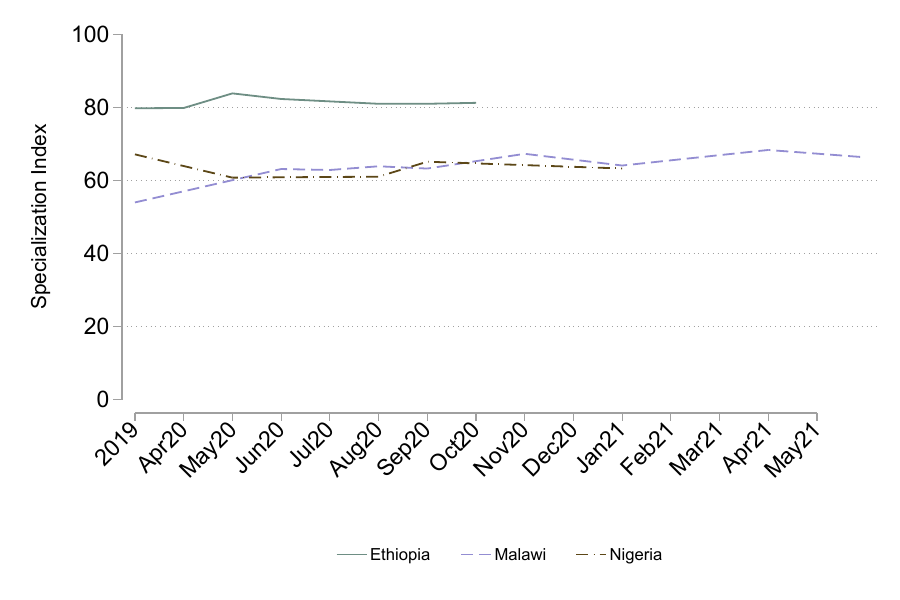}
         \end{center}
        \footnotesize  \textit{Note}: The figure shows the mean values of the fractional index by country over rounds of available data. Higher average values indicate more household specialization (less income diversification). 
    \end{minipage}
\end{figure}

Figure \ref{fig:stp_pp_index_time} illustrates how engagement in income-generating sources changed over time. Recall that our fractional index in constructed so that the number of categories one could earn income from are standardized across country and across time. Despite changes in specific income sources, a fourth stylized fact that emerges from the data is that on average households did not change their income diversification pattern since the onset of the pandemic. We do observe small differences in Ethiopia and Malawi, where households become more specialized during the pandemic than before. These changes are driven by a decline in the percent of households receiving remittances, government assistance, and wage income. This effect is relatively modest. In Nigeria, diversification increased slightly after the start of the pandemic, mainly due to increased participation in farming and greater government assistance.

From these modest stylized facts we conclude that households made limited use of livelihood diversification as a coping strategy after the onset of the pandemic or their attempts to do so were offset by job losses from illness and/or government restrictions throughout 2020.


\subsection{Livelihood Diversification and Food Insecurity}

We present country-level results for our two empirical models, examining the impact of livelihood diversification on household welfare as proxied by food insecurity. Our dynamic panel specification relies on the fractional index measured across all rounds, both pre- and post-outbreak, and aims to test the effectiveness of diversification as an \emph{ex post} coping strategy. Our ANCOVA specification relies on the HHI, which is only measured at baseline, and aims to test the effectiveness of diversification as an \emph{ex ante} coping strategy.

From equation~\ref{eq:dyn}, we are interested in $\beta_1 y_{it-1} + \beta_2 (y_{it-1}*div_{it-1}) + \beta_3 div_{it-1}$. The coefficient $\beta_1$ measures the persistence of food insecurity for the household as current food insecurity is likely to be affected by food insecurity in the previous period. The coefficient $\beta_3$ captures the impact of past decisions about livelihood diversification on current food insecurity, as it takes time for those decision to manifest in income shifts. Finally, the coefficient $\beta_2$ captures the heterogeneity in how past decisions about income diversification impact current food insecurity based on past food insecurity.

Using this framework, we see that by and large our coefficients of interest are relatively small and not significant. Looking first at Table~\ref{tab:dyn_fsi}, coefficients are not significant across all measures of food insecurity across nearly all countries. Of the twelve regressions, the coefficient on the interaction between the lagged fractional index and the lagged food insecurity measure is significant only for the moderate measure of food insecurity in Ethiopia. In Ethiopia, coefficients are small and tend to be negative, suggesting a null or possibly negative effect in which increased specialization may lead to less food insecurity (greater food security). In Malawi and Nigeria, coefficients are small and tend to be positive, suggesting a null or possibly positive effect in which increased specialization may lead to more food insecurity (less food security). However, as 11 of 12 results are null, we do not place great weight on the signs of regression coefficients.

Next, in Table~\ref{tab:inter_fsi} we examine the effects of the triple interaction of the lagged fractional index, lagged food insecurity, and contemporaneous government stringency score. For all four measures of food insecurity and across all three countries, coefficients on the interaction term of interest are not significantly different from zero. That said, the same pattern in signs emerges. Three of four results are small and negative in Ethiopia, potentially suggesting that increased specialization leads to less food insecurity. In Malawi and Nigeria, the opposite pattern holds, potentially suggesting that increased specialization leads to more food insecurity. In this specification, we also observe that COVID stringency is significantly associated with food insecurity, suggesting a positive, albeit small relationship between the strictness of COVID policies and food insecurity, in Ethiopia and Malawi. Based on the results in Tables~\ref{tab:dyn_fsi} and~\ref{tab:inter_fsi}, we conclude that livelihood diversification was not effective as an \emph{ex post} coping strategy for improving food insecurity during the pandemic.

Turning to the ANCOVA specification, Table~\ref{tab:anc_fsi} reports results that relates pre-COVID livelihood diversification, measured using the HHI, to post-outbreak food insecurity measures. Results using ANCOVA are fundamentally similar to those using the dynamic panel regressions. The majority of coefficients on baseline HHI are not significant, with only three coefficients in the interaction term significant. Food insecurity increases in Malawi with more specialization while it decreases in Ethiopia. Coefficients are even closer to zero and confidence intervals are tighter in the ANCOVA regressions, potentially because of the larger sample sizes, since ANCOVA does not rely on a balanced panel of households. We conclude that livelihood diversification was not effective as an \emph{ex ante} coping strategy for improving food insecurity during the pandemic.


\subsection{Interpreting Our Results}

Summarizing the results, the preponderance of evidence points to a lack of a statistically significant relationship between livelihood diversification and food insecurity, our proxy for welfare. We see this both in terms of the stylized facts from our descriptive analysis and in the regression results from our causal analysis. We see households making limited use of livelihood diversification as a coping strategy after the onset of the pandemic. And we see almost no evidence that greater levels of diversification pre-COVID, or changes to diversification during COVID, had a meaningful impact on food insecurity. Based on this, we conclude that the data fails to support our pre-specified hypotheses.

In interpreting our results, there are four potential mechanisms that could offer an explanation. The first is that livelihood diversification truly is effective but that our measure of livelihood diversification or food insecurity does not adequately capture what really matters. This is a reasonable explanation given that there is no single, fully agreed upon way to measure diversification or food insecurity in the literature and no settled theory about if some sources of income matter more than others. Thus, any index of diversification or quantification of food insecurity is inherently \emph{ad hoc}. In anticipation of potential mismeasurement leading to null or unexpected results, we pre-specified a total of six measures of livelihood diversification. In the body of the paper we have presented results from our two preferred diversification measures while in Online Appendix~\ref{sec:data_app} we describe the other four pre-specified indices. As is evident in Online Appendix~\ref{app:main}, a minority of results are statistically significant, with the vast majority of coefficients being not significant. While there are clearly more than just six ways to measure livelihood diversification, we believe the weight of the evidence from these various robustness checks that our main findings are not simply due to mis-categorization of livelihood diversification.

A second explanation for the null findings is that significant results are masked by uncontrolled heterogeneity. Here the logic is that while the average effects are zero, if we conducted a subgroup analysis we would find significant results for these smaller populations. This is exactly what \cite{KhanMorrissey23} find: diversification impacts male versus female headed households differently. Urban households are also differently impacted, relative to rural households. Anticipating the presence of heterogeneous effects, we pre-specified subgroup analysis by gender of the head of household and by whether the household lives in a rural or urban area. Previous research, including research on the effects of COVID-19 \citep{ahmedDeterminantsDynamicsFood2021, josephsonSocioeconomicImpactsCOVID192021, RudinRushetal22}, has show shocks and welfare impacts vary by these sub-populations \citep{mottalebIntraHouseholdResourceAllocation2015, smithDoesResilienceCapacity2018, rahutUnderstandingClimateriskCoping2021}. In Online Appendix~\ref{sec:hetero_app} we present all of our analyses using these pre-specified subgroups. Nearly every estimate is statistically indistinguishable from zero. We find no consistent or coherent evidence that our primary findings of null effects are masking significant effects for certain subgroups.

A third explanation is that our study lacks sufficient power to detect significant effects. The logic here is that we are failing to detect a true, statistically significant, and economically meaningful effect because we lack sufficient observations with sufficient variation. We cannot conclusively rule out this explanation because we did not conduct \emph{ex ante} power calculations as part of our pre-analysis plan. We failed to do this for two reasons. First, it was not clear what values should be used in a power calculation for means and standard deviations in the control group nor what a reasonable expected effect size would be. Second, and quite honestly, we did not think it necessary given the large size of the data collection effort. In total, the pre- and post-outbreak LSMS-ISA data sets contain more than 84,000 observations. Country-specific regressions contain between 3,000 and 15,000 observations. In writing the pre-analysis plan, we did not expect a lack of power to be something we would eventually need to address. Given the evidence on how misleading \emph{ex post} power calculations can be, we have not done this sort of calculation \citep{HoenigHeisey01, Gelman19}. Our failure to conduct \emph{ex ante} power calculations means we cannot provide definitive evidence against the explanation that our analysis lacks power. However, we believe that the large number of observations used in the analysis makes the lack of power an unlikely story.

A final explanation, and the one we find most compelling, is that in the face of a global pandemic and related government restrictions, livelihood diversification was not an effective coping strategy. This is true both in terms of using livelihood diversification as an \emph{ex ante} strategy, to prepare for a potential shock, and an \emph{ex post} strategy, to react to a realized shock. Health concerns and government restrictions to stop the spread of the virus may have stripped resource-rich households of their comparative advantage and equalized vulnerability of income-diverse and income-specialized households. Or perhaps moving away from subsistence farming left households unable to access sufficient food during times of crisis, leaving income-diverse households worse off. In the end, we do not find evidence that income diverse households were better equipped to cope with the socioeconomic impacts of the COVID-19 pandemic. Unlike transitory or localized shocks, which are frequently the setting for research on livelihood diversification as a coping strategy, the pandemic lasted several years and occurred at a global scale. To combat the spread of the virus, governments imposed restrictions on travel and business operations, which may have limited a household's ability to diversify income in response to the pandemic. The evidence presented in this paper suggests that livelihood diversification is ill-fit as a coping strategy for households preparing for or reacting to a shock of the immensity and length of the pandemic.


\section{Conclusions}

The COVID-19 pandemic exacerbated the challenges faced by households in Sub-Saharan Africa. Much of the existing literature suggests that income diversification bolsters household resilience to shocks and improves household welfare after the experience of unanticipated events. However, the focus of this literature is on weather shocks and localized conflict events and so does not capture the nature of a large-scale disaster, like the COVID-19 pandemic. With this study, we seek to fill that gap, exploring two questions related to household income composition and welfare outcomes in this new and unprecedented context. We take advantage of rich survey data to assess trends in income composition over time and to understand the relationship between livelihood diversification and household welfare outcomes, in particular, food security. As our data include a pre-outbreak baseline, we are able to observe household status prior to the pandemic and then track them through and beyond the first year of the pandemic. This panel data, along with our empirical strategies, allows us to identify causal relationships between a household's choice of livelihood activities, income sources, and their level of food insecurity during the pandemic. 

In terms of how household livelihood diversification has changed since the onset of the pandemic, we do not observe substantial or systematic changes in household income composition. Small differences exist in Ethiopia and Malawi, where households become more specialized during the pandemic than before. Conversely, in Nigeria diversification increased slightly after the start of the pandemic. From these trends we conclude that households made limited use of livelihood diversification as a coping strategy during the pandemic.

In terms of how household income composition impacts on food insecurity, our regressions provide little evidence to support the idea that livelihood diversification improved food insecurity during the pandemic. The preponderance of evidence, across countries, estimation methods, and measures of diversification is that there is no significant relationship between livelihood diversification and food insecurity. We provide a number of robustness checks and evidence that we believe demonstrates that the null results are true nulls. Though income diversification may serve as an effective \emph{ex ante} or \emph{ex post} coping mechanism for many shocks, in particular small transitory shocks, we find that for a disaster on the scale of the COVID-19 pandemic this strategy does not appear to be effective. 

An optimistic interpretation of the evidence in this paper would be that the extreme socioeconomic impacts of the pandemic appear to necessitate alternative adaptation strategies. A pessimistic interpretation is that a pandemic is too disastrous and omnipresent to prepare for or adequately adapt to. Either interpretation leads to the conclusion that commonly promoted coping strategies, such as livelihood diversification, that households are encouraged to undertake on their own are inadequate for large-scale or long-term disruptions and disasters. As households, development agencies, and governments look to prepare for the increased occurrence of such disasters, either due to climate change or the increased spread of zoonotic disease, this point bears keeping in mind. Future research and development will need to grapple with the fact that the coping strategies that gave people hope in the past may fail them as they try to cope with the increased scale of shocks in the future.


\newpage
\singlespacing
\bibliographystyle{chicago}
\bibliography{references}


\newpage 
\FloatBarrier

\pagebreak
{\footnotesize
\input{./tables/inc_sum}
}
\restoregeometry

\begin{landscape}
{\small
\input{./tables/indices_sum_main}
}
\end{landscape}

\begin{landscape}
\begin{table}[htbp]	\centering
    \caption{Dynamic Panel Regressions} \label{tab:dyn_fsi}
	\scalebox{.75}
	{ \setlength{\linewidth}{.2cm}\newcommand{\contents}
		{\input{./tables/dyn_fs}}
	\setbox0=\hbox{\input{./tables/inc_bond_fs}}
    \setlength{\linewidth}{\wd0-2\tabcolsep-.25em}
    \input{./tables/inc_bond_fs}}
\end{table}
\end{landscape}

\begin{landscape}
\begin{table}[htbp]	\centering
    \caption{Dynamic Panel Regressions with Government Stringency Score} \label{tab:inter_fsi}
	\scalebox{.65}
	{ \setlength{\linewidth}{.2cm}\newcommand{\contents}
		{\input{./tables/inter_fs}}
	\setbox0=\hbox{\input{./tables/inc_bond_fs}}
    \setlength{\linewidth}{\wd0-2\tabcolsep-.25em}
    \input{./tables/inc_bond_fs}}
\end{table}
\end{landscape}

\begin{landscape}
\begin{table}[htbp]	\centering
    \caption{ANCOVA Regressions} \label{tab:anc_fsi}
	\scalebox{.75}
	{ \setlength{\linewidth}{.2cm}\newcommand{\contents}
		{\input{./tables/anc_fs}}
	\setbox0=\hbox{\input{./tables/inc_bond_fs}}
    \setlength{\linewidth}{\wd0-2\tabcolsep-.25em}
    \input{./tables/inc_bond_fs}}
\end{table}
\end{landscape}


\clearpage
\newpage
\appendix
\onehalfspacing

\begin{center}
	\section*{Online-Only Online Appendix to ``Coping or Hoping? Livelihood Diversification and Household Welfare in the COVID-19 Pandemic''} \label{sec:app}
\end{center}


\section{Additional Data Considerations} \label{sec:data_app}

\setcounter{table}{0}
\renewcommand{\thetable}{A\arabic{table}}
\setcounter{figure}{0}
\renewcommand{\thefigure}{A\arabic{figure}}

\subsection{Diversification Indices} \label{sec:div_app}

Following our pre-analysis plan, in addition to the indices specified and discussed in the main text, we also present findings from four additional indices. These are summarized in Table~\ref{tab:indices_app}.  

Each of the indices, here and in the paper, has advantages and limitations. The fractional indices consider engagement in income-generating activities. These measures include binary responses and the dichotomous nature of these variables allows for comparison in income-generating activities over time with the inclusion of the post-outbreak HFPS rounds. However, these fractional indices do not consider the amount of income earned from each source. As such, these indices are a less nuanced representation of household income diversity than the Herfindahl-Hirschman Index (HHI) measures. For example, suppose Household A was engaged in casual employment for one week in 2019. During that week, the household earned five percent of their total annual income and the remaining 95 percent was generated through farm work. Suppose their neighbor, Household B, was also engaged in casual labor and farm work but generated equal incomes from these two sources (a 50 percent split). In our data, Households A and B would receive the same fraction score, even though Household A is much more dependent on a single income source than Household B and as a result Household A would have a higher HHI than Household B. There is a trade-off between using all of the data, pre- and post-outbreak, and just the pre-COVID data. The former gives us more observations over time but less detail about income. The later provides more detail about income but is only a snap shot in time. 

Similarly, there is a trade-off between the simplicity of the fractional indices and the HHI. The fractional indices lack detail because they encode that detail in simple ``yes'' or ``no'' answers. Conversely, the HHI indices consider the portion of total income generated from each source, providing a more detailed measure of income diversity. However, these values are influenced by outliers in the data. Income calculations often involve multiplication of different variables:

\begin{align*}
    wage \; earnings = \; & hourly\;income\; *\; hours\;worked\;per\; week\; * \\
     & weeks\; worked\; per\; month\; *\; months\; worked\; per\; year,
\end{align*}

\noindent aggregation across income sub-categories:

\begin{align*}
    income\; from\; livestock\; products\; &= income\; from\; milk\; sales\; +\; value\; of\; household\; milk\; \\
    &consumption\; +\;  income from\; meat\; sales\; +\; value\; of\; household\; \\
    & meat\; consumption + ... ,
\end{align*}

\noindent and other data manipulations. An error, misrepresentation, or miscalculation in any one of these intermediate variables can lead to erroneous estimations, which compound as one continues to aggregate values. Additionally, when considering crop and livestock product income, prices are not available for household consumption. Following standard practices in the literature, we assume the value of a consumed product is equal to the median sale price for that product in the household's geographic area. To account for large outliers, for each income category we winsorize outliers greater than two standard deviations from the median and impute their values. Despite this adjustment, the data are still vulnerable to potential error and subjective assumptions that affect their accuracy, which may lead to inaccuracy and/or bias in our estimated values. Because HHI scores are calculated based on a percentage, a measurement inaccuracy in one income source can distort the overall score. 

\subsection{Food Insecurity Experience Scale Questions} \label{sec:fies_app}

Tables \ref{tab:fies_qs_eth} through \ref{tab:fies_qs_nga} shows survey questions used to measure food insecurity in each country included in this study.


\begin{landscape}
\newgeometry{left=2cm, top=10.3cm}
\pagestyle{empty}
{\small
\input{./tables/indices_sum_app}
}
\restoregeometry
\end{landscape}

\begin{landscape}
\begin{table}[htbp]	\centering
    \caption{Food Insecurity Experience Scale (FIES) Questions in Ethiopia} \label{tab:fies_qs_eth}
	\scalebox{.9}
	{ \setlength{\linewidth}{.1cm}\newcommand{\contents}
		{\begin{tabular}{c p{0.95\linewidth} c p{0.95\linewidth} c}
            \\[-1.8ex]\hline 
			\hline \\[-1.8ex]
            & \multicolumn{2}{c}{Pre-COVID-19 Surveys} & \multicolumn{2}{c}{COVID-19 Surveys} \\[1ex]
            FIES & Question & Recall & Question & Recall \\
            \midrule
            FS1 & Did you worry that your household would not have enough food? & 7 day & Was there a time when you or any other adult in your household were worried about not having enough food to eat because of lack of money or other resources? & 30 day \\
            FS2 & How many days have you or someone in your household had to rely on less preferred foods? & 7 day & Was there a time when you, or any other adult in your household, were unable to eat healthy and nutritious/preferred foods because of a lack of money or other resources? & 30 day \\
            FS3 & How many days have you or someone in your household had to limit the variety of foods eaten? & 7 day & Was there a time when you, or any other adult in your household, ate only a few kinds of foods because of a lack of money or other resources? & 30 day \\
            FS4 & How many days have you or someone in your household had to limit portion size at meal-times? & 7 day & Was there a time when you or others in your household had to skip a meal because there was not enough money or other resources to get food? & 30 day \\
            FS5 & How many days have you or someone in your household had to reduce number of meals eaten in a day? & 7 day & Was there a time when you or others in your household ate less than you thought you should because of a lack of money or other resources? & 30 day \\
            FS6 & How many days have you had no food of any kind in your household? & 7 day & Was there a time when your household ran out of food because of a lack of money or other resources? & 30 day \\
            FS7 & How many days have you or someone in your household had to restrict consumption by adults for small children to eat? & 7 day & Was there a time when you or others in your household were hungry but did not eat because there was not enough money or other resources for food? & 30 day \\
            FS8 & How many days have you or someone in your household had to go a whole day and night without eating anything? & 7 day & Was there a time when you or others in your household went without eating for a whole day because of a lack of money or other resources? & 30 day \\
			\\[-1.8ex]\hline 
			\hline \\[-1.8ex]
    		
    	\end{tabular}}
	\setbox0=\hbox{\input{./tables/inc_bond_fs}}
    \setlength{\linewidth}{\wd0-2\tabcolsep-.25em}
    \input{./tables/inc_bond_fs}}
\end{table}
\end{landscape}

\begin{landscape}
\begin{table}[htbp]	\centering
    \caption{Food Insecurity Experience Scale (FIES) Questions in Malawi} \label{tab:fies_qs_mwi}
	\scalebox{.9}
	{ \setlength{\linewidth}{.1cm}\newcommand{\contents}
		{\begin{tabular}{c p{0.95\linewidth} c p{0.95\linewidth} c}
            \\[-1.8ex]\hline 
			\hline \\[-1.8ex]
            & \multicolumn{2}{c}{Pre-COVID-19 Surveys} & \multicolumn{2}{c}{COVID-19 Surveys} \\[1ex]
            FIES & Question & Recall & Question & Recall \\
            \midrule
            FS1 & Did you worry that your household would not have enough food? & 7 day & You or any other adult in your household were worried about not having enough food to eat because of lack of money or other resources? & 30 day \\
            FS2 & How many days have you or someone in your household had to rely on less preferred and/or less expensive foods?& 7 day & You, or any other adult in your household, were unable to eat healthy and nutritious/preferred foods because of a lack of money or other resources? & 30 day \\
            FS3 & & 7 day & You, or any other adult in your household, ate only a few kinds of foods because of a lack of money or other resources? & 30 day \\
            FS4 & How many days have you or someone in your household had to reduce number of meals eaten in a day? & 7 day & You, or any other adult in your household, had to skip a meal because there was not enough money or other resources to get food? & 30 day \\
            FS5 & How many days have you or someone in your household had to limit portion size at meal-times? & 7 day & You, or any other adult in your household, ate less than you thought you should because of a lack of money or other resources? & 30 day \\
            FS6 & How many days have you or someone in your household had to borrow food or rely on help from a friend or relative? & 7 day & Your household ran out of food because of a lack of money or other resources? & 30 day \\
            FS7 & How many days have you or someone in your household had to restrict consumption by adults in order for small children to eat? & 7 day & You, or any other adult in your household, were hungry but did not eat because there was not enough money or other resources for food? & 30 day \\
            FS8 & & 7 day & You, or any other adult in your household, went without eating for a whole day because of a lack of money or other resources? & 30 day \\
			\\[-1.8ex]\hline 
			\hline \\[-1.8ex]
    		
    	\end{tabular}}
	\setbox0=\hbox{\input{./tables/inc_bond_fs}}
    \setlength{\linewidth}{\wd0-2\tabcolsep-.25em}
    \input{./tables/inc_bond_fs}}
\end{table}
\end{landscape}
  
\begin{landscape}
\begin{table}[htbp]	\centering
    \caption{Food Insecurity Experience Scale (FIES) Questions in Nigeria} \label{tab:fies_qs_nga}
	\scalebox{.9}
	{ \setlength{\linewidth}{.1cm}\newcommand{\contents}
		{\begin{tabular}{c p{0.95\linewidth} c p{0.95\linewidth} c}
            \\[-1.8ex]\hline 
			\hline \\[-1.8ex]
            & \multicolumn{2}{c}{Pre-COVID-19 Surveys} & \multicolumn{2}{c}{COVID-19 Surveys} \\[1ex]
            FIES & Question & Recall & Question & Recall \\
            \midrule
            FS1 & You or any other adult in your household worried about not having enough food to eat because of lack 
of money or other resources? & 30 day & You or any other adult in your household were worried about not having enough food to eat because of lack of money or other resources? & 30 day \\
            FS2 & You, or any other adult in your household, were unable to eat healthy and nutritious/preferred foods because of a lack of money or other resources? & 30 day & You, or any other adult in your household, were unable to eat healthy and nutritious/preferred foods because of a lack of money or other resources? & 30 day \\
            FS3 & You, or any other adult in your household, ate only a few kinds of foods because of a lack of money or other resources? & 30 day & You, or any other adult in your household, ate only a few kinds of foods because of a lack of money or other resources? & 30 day \\
            FS4 & You, or any other adult in your household, had to skip a meal because there was not enough money or other resources to get food? & 30 day & You, or any other adult in your household, had to skip a meal because there was not enough money or other resources to get food? & 30 day \\
            FS5 & You, or any other adult in your household, restricted consumption in order for children to eat? & 30 day & You, or any other adult in your household, ate less than you thought you should because of a lack of money or other resources? & 30 day \\
            FS6 & You, or any other adult in your household, borrowed food, or relied on help from a friend or relative? & 30 day & Your household ran out of food because of a lack of money or other resources? & 30 day \\
            FS7 & You, or any other adult in your household, restricted consumption in order for children to eat? & 30 day & You, or any other adult in your household, were hungry but did not eat because there was not enough money or other resources for food? & 30 day \\
            FS8 & You, or any other adult in your household, went without eating for a whole day because of a lack of money or other resources? & 30 day & You, or any other adult in your household, went without eating for a whole day because of a lack of money or other resources? & 30 day \\
			\\[-1.8ex]\hline 
			\hline \\[-1.8ex]
    		
    	\end{tabular}}
	\setbox0=\hbox{\input{./tables/inc_bond_fs}}
    \setlength{\linewidth}{\wd0-2\tabcolsep-.25em}
    \input{./tables/inc_bond_fs}}
\end{table}
\end{landscape}


\clearpage
\newpage
\onehalfspacing

\section{Robustness Checks on Main Analysis} \label{app:main}

\setcounter{table}{0}
\renewcommand{\thetable}{B\arabic{table}}
\setcounter{figure}{0}
\renewcommand{\thefigure}{B\arabic{figure}}

\subsection{Results Using Alternative Indices} \label{sec:alt_indx_app}

Per our pre-analysis plan, we pre-specified tests of our primary hypotheses using six indices. After having collected and cleaned the data and conducted the analysis, we found that nearly all estimates of the coefficients of interest were not statistically significant. Because of this, we simplified the presentation of our findings in the main body of the paper to rely solely on our two preferred indices. These indices are the fractional index and HHI that standardize income categories over country and over time. This allows for the cleanest possible comparison between countries and pre/post-outbreak. To complete the pre-specified analysis, we present the results from all of our main empirical specifications using the other four indices. In short, results do not differ when using any of these other four pre-specified indices.

Table~\ref{tab:index2} presents results using the what we term Fractional Index 2. This index is similar to our preferred fractional index in that is standardizes income categories across time (pre- and post-outbreak). It differs from our preferred index in that it does not standardize income categories across country. This means that each country can have a different number of income sources, taking full advantage of the richness of the data but making cross country comparisons more difficult. Results using fractional index 2 are not meaningfully different from those with our preferred fractional index. Nearly all coefficients of interest are not significant.

Table~\ref{tab:app_anc_fs} provides results for what we terms fractional indices 3 and 4 and HHI 2. Fractional index 3 standardizes income categories across countries but uses only the pre-COVID data for construction. Fractional index 4 also uses only pre-COVID data but does not standardize across countries. Recall our preferred fractional index standardizes across both time and country, making the index comparable across these dimensions but also resulting in the fewest categories. Fractional index 4 uses all available data, resulting in the most income categories possible for each country, but the loss of standardization means the index values from one country are not comparable to another. Like fractional index 4, income categories in HHI 2 are not standardized across country.

In Tables~\ref{tab:app_anc_fs} and~\ref{tab:app_did_fs} we present results of all three indices' impact on food security using the ANCOVA and difference-in-difference specifications. Note that since all of these indices use only baseline data we cannot employ them in the dynamic panel models. Similar to the results using our two preferred indices, that vast majority of point estimates are not statistically significant.

We conclude that our primary results, which show no significant relationship between livelihood diversification and welfare outcomes, are not an artifact of our definition of livelihood diversification. Using alternative, pre-specified indices does not change our findings.

In this Online Appendix we present additional information on our main results. First, we present tabular versions of the main results from the paper. Second, we present results from our pre-specified difference-in-difference regressions, which serve as a robustness check on our preferred ANCOVA results.

In terms of our main results, Table~\ref{tab:inter_fsi} corresponds to a figure in the main text. The graphical representation of results in the main text is limited to our variable of interest: an interaction term in the dynamic panel model and the value of the pre-COVID index in the ANCOVA model. While the coefficient plots are succinct and condensed, they lack information on sample size as well as coefficient estimates on other terms that might be of interest in the regressions. Below we present results from our main specifications in tabular form so as to provide more complete information for the interested reader.

\subsection{Results Using Alternative Specifications} \label{app_did}

In addition to ANCOVA specifications, we estimate a simple difference-in-difference model in which we include an indicator for the start of COVID-19-related restrictions in Sub-Saharan Africa. This specification takes on the following functional form: 

\begin{equation}\label{eq:did}
    y_{it}=\alpha+\beta_1div_{it=0}+\beta_2 \left( div_{it=0}*covid_t \right) +\beta_3covid_t+\delta_t+c_j*t_t+u_i+\epsilon_{it}.
\end{equation}

Here $div_{it=0}$ is the diversity index in the pre-pandemic period and $covid_t$ is an indicator for before and after the start of the pandemic. The variable of interest in this specification is $\beta_2$, the difference-in-difference effect of income diversity post-pandemic.

\begin{figure}[htbp]
    \begin{minipage}{\linewidth}
         \caption{Difference-in-Difference Regressions [Dependent Variable: Food Insecurity]} \label{fig:fs_did}
        \begin{center}
            \includegraphics[width=\linewidth,keepaspectratio]{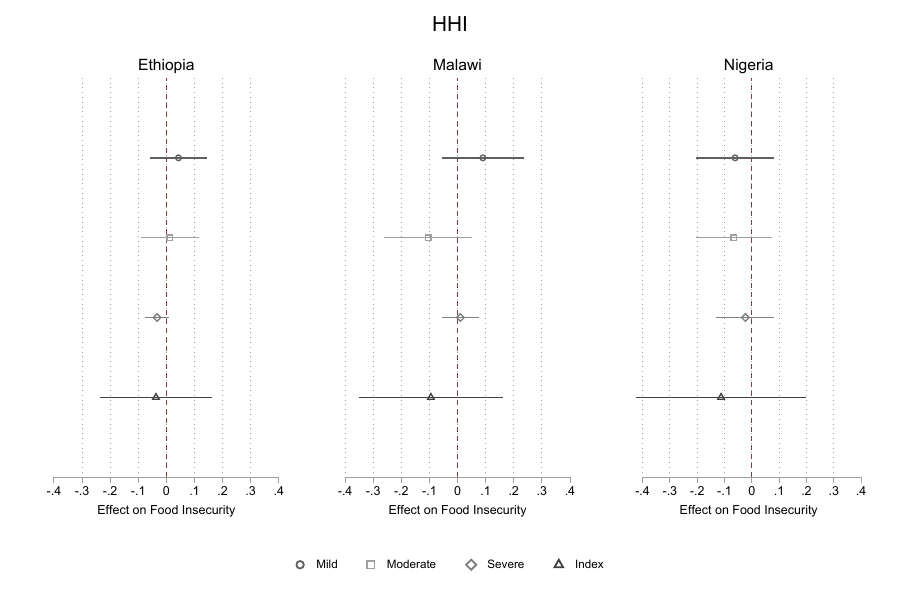}
         \end{center}
        \footnotesize \textit{Note}: The figure plots difference-in-difference regression results with region and round controls (see Equation \ref{eq:did}). We display coefficients for baseline, pre-pandemic, Herfindahl-Hirschman Index (HHI) for Ethiopia, Malawi, and Nigeria. Horizontal lines represent 95 percent confidence intervals. Horizontal lines represent 95 percent confidence intervals based on standard errors clustered at the household level.
    \end{minipage}
\end{figure}

As seen in Figure~\ref{fig:fs_did}, results for food insecurity from the difference-in-difference specifications are generally consistent with their ANCOVA counterpart (Table~\ref{tab:anc_fsi}) but with less precise measures of standard errors. The same is true for results on educational engagement.

While not pre-specified, we also test the robustness of our results to using using the Arellano-Blover/Blundell-Bond \citep{ArellanoBover95, BlundellBond98} GMM estimator. These results are presented in Table~\ref{tab:inc_bondfs}. With this specification, we demonstrate that the size of the downward bias is small and does not affect the interpretation of our results, with both specifications mirroring one another. 


\begin{landscape}
\begin{table}[htbp]
    \centering
    \footnotesize
    \caption{Dynamic Panel \& Panel with Interactions Regressions - Fractional Index 2 [Dependent Variable: Food Insecurity]} \label{tab:index2}
	\scalebox{.93}
		{\input{./tables/app_index2}}
\end{table}
\end{landscape}

\begin{landscape}
\begin{table}[htbp]
    \centering
    \footnotesize
    \caption{ANCOVA Regressions [Dependent Variable: Food Insecurity]} \label{tab:app_anc_fs}
	\scalebox{.95}
		{\input{./tables/app_anc_fs}}
\end{table}
\end{landscape}


\newpage 

\begin{landscape}
\begin{table}[htbp]
    \centering
    \footnotesize
    \caption{Difference-in-Difference Regressions [Dependent Variable: Food Insecurity]} \label{tab:app_did_fs}
	\scalebox{.95}
		{\input{./tables/app_did_fs}}
\end{table}
\end{landscape}

\begin{table}[htbp]	\centering
    \caption{Arellano–Bond Dynamic Panel Estimation} \label{tab:inc_bondfs}
	\scalebox{.8}
	{ \setlength{\linewidth}{.2cm}\newcommand{\contents}
		{\input{./tables/inc_bond_fs}}
	\setbox0=\hbox{\input{./tables/inc_bond_fs}}
    \setlength{\linewidth}{\wd0-2\tabcolsep-.25em}
    \input{./tables/inc_bond_fs}}
\end{table}

\clearpage
\newpage
\onehalfspacing

\section{Heterogeneous Effects Analysis} \label{sec:hetero_app}

\setcounter{table}{0}
\renewcommand{\thetable}{D\arabic{table}}
\setcounter{figure}{0}
\renewcommand{\thefigure}{D\arabic{figure}}

In our pre-analysis plan we proposed to investigate the heterogeneous effects of livelihood diversification for different population subgroups. Specifically, we planned to assess differences for male- and female-headed households as well as urban and rural households. We did not present these results in the main body of the paper since our primary results were not significant and because the subgroup analysis also produced null results. For completeness, we discuss the method used and results of the sub-group analysis here.

In Table \ref{tab:ind_sec_sex} we present the distributions of all of the indices comparing urban and rural as well as male- and female-headed households. We only include the pre-COVID-19 data, even when post-outbreak rounds are available for that index. Urban households tend to be more specialized than rural households, a result that is particularly evident in Ethiopia. There are not notable differences in income diversification by head-of-household gender. 

\subsection{Method}

To estimate heterogeneous effects of livelihood diversification on welfare outcomes use the ANCOVA specifications discussed in section~\ref{sec:method}. We interact the sub-group indicator variables with livelihood diversification at baseline to understand the differential impacts for female headed households and rural households. 

\begin{equation}\label{eq:anc_app}
 y_{it}=\alpha+\beta_1div_{it=0}+\beta_2 \left( div_{it=0}*sub_{i} \right) +\beta_3sub_{i}+ \beta_4y_{it=0}+\delta_t+r_j*t_t+\epsilon_{it}   
\end{equation}

\noindent $sub_{i}$ is an indicator variable for population subgroups based on head-of-household gender or household sector for household $i$. All other terms are as previously defined. The interaction term, $\beta_2$, represents the differential impact of pre-COVID-19 livelihood diversification on household welfare outcomes for these population subgroups. Standard errors are clustered at the household level.

Similarly, we estimate heterogeneous effects using a standard difference-in-difference model, with out coefficient of interesting being on the triple difference effect of COVID-diversification-female headed/rural household $(\beta_7)$:

\begin{align}\label{eq:did_app}
    y_{it} &= \alpha + \beta_1 div_{it=0} + \beta_2 \left( div_{it=0}*covid_t \right) + \beta_3 covid_t + \beta_4 \left( div_{it=0}*sub_{i} \right) + \beta_5 \left( covid_t*sub_{i} \right)  \nonumber \\
    &+ \beta_6 sub_{i} + \beta_7 + \left( div_{it=0}**covid_t*sub_{i} \right) + \delta_t + c_j*t_t + \epsilon_{it}.
\end{align}

\noindent All other terms are previously defined and standard errors are clustered at the household.

\subsection{Heterogeneous Effects of Livelihood Diversification}

In the main paper we address two pre-specified research questions. In this section, we explore our third pre-specified research question: does income diversification have disparate impacts on different country subgroups in the context of the COVID-19 pandemic? Specifically, we investigate heterogeneous impacts for male- and female-headed households as well as urban and rural households. To detect these potentially disparate effects, we include binary interactions term in our ANCOVA and difference-in-difference specifications indicating head-of-household gender and household sector. Because we use only the ANCOVA and difference-in-difference specification to answer this question, we restrict our analysis to just the indices that rely on baseline data (fractional indices 3 and 4 and HHI 1 and 2).

Table~\ref{tab:fs_anc_sex} displays coefficient estimates for the ANCOVA specification with a head-of-household gender interaction. In this context, male-headed households are the comparison group. While nearly every coefficient of interest is not significant, coefficients tend to be slightly positive coefficients, suggesting that households headed by women may experience increased food insecurity when household incomes are more specialized. As seen in Table~\ref{tab:fs_did_sex}, using the difference-in-difference estimator also produces null results with a similar tendency for most coefficients to be slightly positive. 

We also test for heterogeneous impacts across rural and urban populations. Similar to the results for differences based on gender of the head-of-household, all of the results for differences between urban and rural households are statistically insignificant. Unlike the head-of-household gender results, the specifications that include urban/rural indicators do not point to a consistent relationship in terms of sign. For these specifications, rural households serve as the comparison group.

As seen in Table~\ref{tab:fs_anc_sec} and Table~\ref{tab:fs_did_sec}, coefficients for the interaction term do not evidence a differential impact of livelihood diversification on food security for urban versus rural populations. Coefficient estimates are never statistically significant and do not follow a discernible trend across countries.

Overall we do not find any significant differences in how livelihood diversification impacts welfare outcomes based on the gender of the head-of-household or whether the household is rural or urban. There is a slight pattern of female headed households experiencing worse outcomes than male headed households when female headed households are more specialized. No pattern emerges for urban/rural households.
 

\pagebreak
{\footnotesize
\input{./tables/ind_density_sec_sex}
}
\restoregeometry


\begin{landscape}
\begin{table}[htbp] 
    \centering
    \footnotesize
    \caption{ANCOVA Regressions by Head-of-Household Gender [Dependent Variable: Food Insecurity]} \label{tab:fs_anc_sex}
	\scalebox{1}
		{\input{./tables/fs_anc_sex}}
\end{table}
\end{landscape}

\begin{landscape}
\begin{table}[htbp] 
    \centering
    \footnotesize
    \caption{Difference-in-Difference Regressions by Head-of-Household Gender [Dependent Variable: Food Insecurity]} \label{tab:fs_did_sex}
	\scalebox{1}
		{\input{./tables/fs_did_sex}}
\end{table}
\end{landscape}


\begin{landscape}
\begin{table}[htbp]
    \centering
    \footnotesize
    \caption{ANCOVA Regressions by Urban/Rural [Dependent Variable: Food Insecurity]} \label{tab:fs_anc_sec}
	\scalebox{1}
		{\input{./tables/fs_anc_sec}}
\end{table}
\end{landscape}

\begin{landscape}
\begin{table}[htbp] 
    \centering
    \footnotesize
    \caption{Difference-in-Difference Regressions by Urban/Rural [Dependent Variable: Food Insecurity]} \label{tab:fs_did_sec}
	\scalebox{1}
		{\input{./tables/fs_did_sec}}
\end{table}
\end{landscape}

\end{document}

%% file: tables/inc_sum.tex
\begin{longtable}{l ccc} 
\caption{Pre-COVID Engagement in and Earnings from Income Sources} \label{incsum} \\ [-1.8ex] \hline \hline 
& & \multicolumn{2}{c}{\textbf{Income (USD)}} \\  
& \multicolumn{1}{c}{\textbf{Share Engaged}} & \multicolumn{1}{c}{\textbf{Mean}} & \multicolumn{1}{c}{\textbf{Standard Deviation}} \\  
\endfirsthead 
\hline & & \multicolumn{2}{c}{\textbf{Income (USD)}} \\  
& \multicolumn{1}{c}{\textbf{Share Engaged}} & \multicolumn{1}{c}{\textbf{Mean}} & \multicolumn{1}{c}{\textbf{Standard Deviation}} \\ \midrule 
\endhead 
\midrule \multicolumn{4}{c}{{Continued on Next Page\ldots}} 
\endfoot 
\endlastfoot 
\midrule \multicolumn{4}{c}{\textit{Panel A: Ethiopia}} \\ 
\midrule
Crop Income & 0.522 &      279 &      498 \\ 
Livestock Sales & 0.318 &      249 &      240 \\ 
Livestock Product Income & 0.489 &      533 &    1,202 \\ 
Wages & 0.215 &    1,490 &    3,740 \\ 
Casual Employment Wages & 0.088 &      179 &      387 \\ 
Temporary Employment Wages & 0.087 &       96 &       97 \\ 
Non-Farm Enterprises & 0.224 &    1,271 &    2,933 \\ 
In-Kind Transfers/Gifts & 0.024 &       57 &       96 \\ 
Cash Transfers/Gifts & 0.099 &      304 &      504 \\ 
Food Transfers/Gifts & 0.048 &       61 &       79 \\ 
In-kind Transfers from Govt and NGOs & 0.009 &       39 &       30 \\ 
Cash Transfers from Govt and NGOs & 0.032 &       64 &       62 \\ 
Free Food & 0.049 &       36 &       47 \\ 
Pension & 0.013 &      279 &      242 \\ 
Rental Income & 0.084 &      337 &      584 \\ 
Asset Sales & 0.087 &      262 &      244 \\ 
Savings, Interest, Investment & 0.002 &      145 &      355 \\ 
Other & 0.008 &      354 &      424 \\ 
\midrule
\multicolumn{1}{l}{Observations} & 
\multicolumn{3}{c}{3,247} \\ \midrule   
\multicolumn{4}{c}{\textit{Panel B: Malawi }} \\ 
\midrule
Crop Income & 0.776 &      131 &      206 \\ 
Tree Crop Sales & 0.062 &       28 &       32 \\ 
Livestock Sales & 0.263 &       56 &       76 \\ 
Livestock Product Income & 0.298 &       42 &       89 \\ 
Wages & 0.261 &    1,617 &    3,109 \\ 
Casual Employment Wages & 0.609 &      314 &      453 \\ 
Non-Farm Enterprises & 0.438 &    1,894 &    5,154 \\ 
Cash Transfers/Gifts & 0.263 &       68 &      131 \\ 
Food Transfers/Gifts & 0.272 &       13 &       11 \\ 
In-Kind Transfers/Gifts & 0.114 &       30 &       50 \\ 
Cash from Children & 0.197 &       65 &       87 \\ 
In-Kind Transfers from children & 0.131 &       45 &       48 \\ 
Free Food & 0.182 &       20 &       14 \\ 
Cash Transfers from Govt and NGOs & 0.064 &       57 &       36 \\ 
Cash or Inputs for Work & 0.019 &       56 &       28 \\ 
MASAF Public Works Program & 0.043 &       32 &       15 \\ 
Pension & 0.011 &    1,355 &    1,320 \\ 
Rental Income & 0.080 &      321 &      392 \\ 
Asset Sales & 0.076 &       87 &      102 \\ 
Savings, Interest, Investment & 0.068 &       64 &       77 \\ 
Other & 0.043 &       43 &      108 \\ 
\midrule
\multicolumn{1}{l}{Observations} & 
\multicolumn{3}{c}{1,726} \\ \midrule   
\multicolumn{4}{c}{\textit{Panel C: Nigeria }} \\ 
\midrule
Crop Income & 0.644 &      401 &      449 \\ 
Tree Crop Sales & 0.064 &      233 &      369 \\ 
Livestock Sales & 0.216 &      165 &      214 \\ 
Livestock Product Income & 0.187 &       93 &      178 \\ 
Wages & 0.260 &    1,598 &    1,419 \\ 
Non-Farm Enterprises & 0.624 &    2,926 &    4,109 \\ 
Domestic Remittances & 0.266 &      110 &      115 \\ 
Foreign Remittances & 0.034 &      267 &      295 \\ 
In-Kind Remittances & 0.125 &       47 &       61 \\ 
Cash, Food, or In-kind Assistance & 0.042 &       54 &       54 \\ 
Pension & 0.030 &      689 &      900 \\ 
Rental Income (Non-Ag) & 0.049 &      347 &      437 \\ 
Rental Income (Ag) & 0.039 &       45 &       88 \\ 
Savings, Interest, Investment & 0.021 &      277 &      804 \\ 
Other & 0.011 &      610 &      485 \\ 
\midrule
\multicolumn{1}{l}{Observations} & 
\multicolumn{3}{c}{1,950} \\ \midrule \multicolumn{4}{p{360pt}}{\footnotesize \textit{Note}: The table displays the share of households engaged in each category of livelihood activity and the mean and standard deviation of income earned from that category. In the LSMS-ISA data, income is reported in the local currency. To allow for cross-country comparisons, we convert income values to US dollars using 2019 exchange rates.}  \\ \end{longtable}  

%% file: tables/indices_sum_main.tex
\begin{longtable}{P{1cm} P{2cm} P{1.5cm} P{8.5cm} P{8.5cm}} \caption{Livelihood Diversification Indices Summary}   \label{tab:indices_main} \\ [-1.8ex]\hline \hline

Index Type &  Standardized Across Countries & Time Period & Description & Pre-COVID-19 Kernel Density Graph \\ \hline
\centering
Fraction & 	Yes & 	Pre- and Post-COVID-19 &
To generate this fraction index, we collapse multiple income sources into seven broad income-generation categories that are consistent across rounds and across countries. These categories are: farm; wage; pension; remittances; non-farm enterprise; income from properties, investments and savings; and other. The ``other'' income category varies across countries and rounds but generally includes asset sales, income from NGOs, and government assistance. & 
    \includegraphics[width=85mm, scale=1]{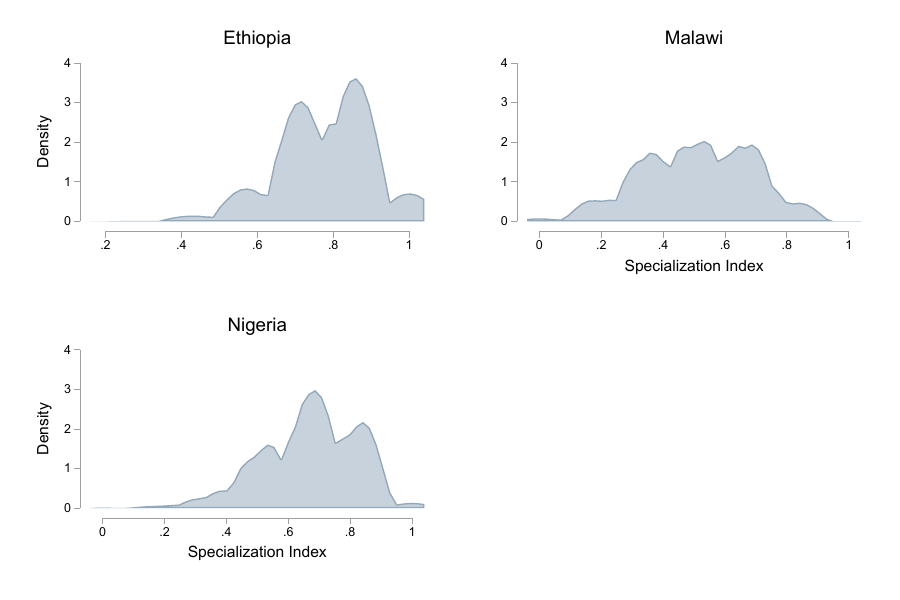} \\
\midrule 
HHI	& Yes &	Pre-COVID-19 &	Given the level of detail provided in the pre-COVID survey data, we are able to generate an HHI to capture household income diversity more precisely. For this index, we use the same 12 income categories used in the simple fraction index but consider the amount earned from each source. & 
    \includegraphics[width=85mm, scale=1]{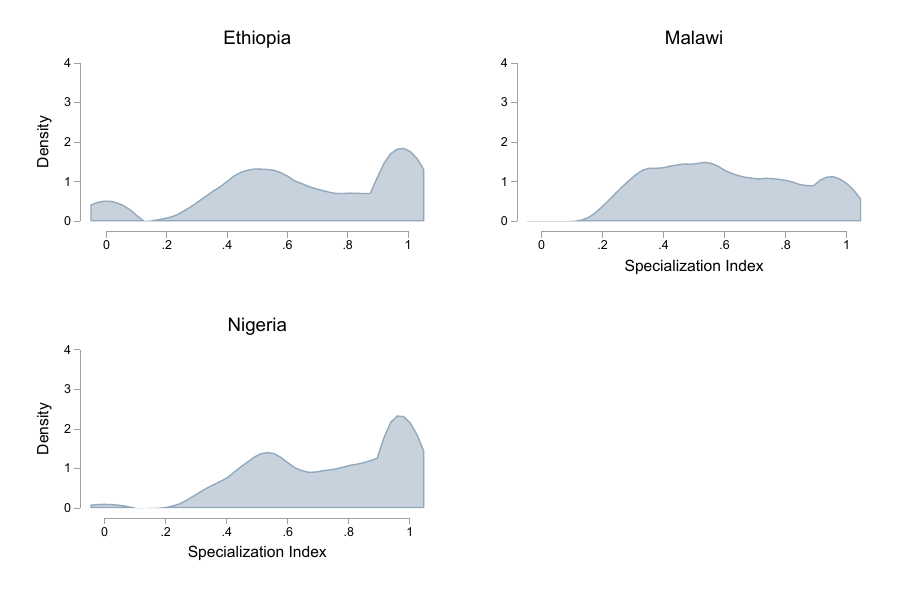} \\ [-1.8ex]\hline \hline
\multicolumn{5}{p{23cm}}{\footnotesize \textit{Note}: The table summarizes the two livelihood diversification indices used in the main analysis. Higher index values indicate more household specialization (less income diversification). Appendix~\ref{sec:data_app} contains similar summary information regarding the other four indices we pre-specified.}
\end{longtable}

%% file: tables/dyn_fs.tex
\begin{tabular}{l*{12}{c}} \\ [-1.8ex]\hline \hline \\[-1.8ex] 
& \multicolumn{4}{c}{Ethiopia} & \multicolumn{4}{c}{Malawi} & 
\multicolumn{4}{c}{Nigeria} \\ & \multicolumn{1}{c}{FS Index} & \multicolumn{1}{c}{Mild} & 
\multicolumn{1}{c}{Moderate} & \multicolumn{1}{c}{Severe} & \multicolumn{1}{c}{FS Index} 
& \multicolumn{1}{c}{Mild} & \multicolumn{1}{c}{Moderate} & \multicolumn{1}{c}{Severe} 
& \multicolumn{1}{c}{FS Index} & \multicolumn{1}{c}{Mild} & \multicolumn{1}{c}{Moderate} 
& \multicolumn{1}{c}{Mild} \\ \midrule 
\midrule
Lagged Fractional Index (FI)&      -0.161         &       0.009         &       0.027         &      -0.105         &       0.139         &      -0.052         &       0.044         &      -0.031         &       0.309         &       0.028         &       0.020         &       0.071         \\
                    &     (0.131)         &     (0.126)         &     (0.080)         &     (0.054)         &     (0.090)         &     (0.087)         &     (0.071)         &     (0.043)         &     (0.192)         &     (0.156)         &     (0.117)         &     (0.075)         \\
Lagged FS $\times$ Lagged FI&      -0.180         &      -0.086         &      -0.339\sym{*}  &       0.525         &      -0.003         &       0.129         &       0.032         &       0.060         &       0.165         &      -0.005         &       0.215         &       0.087         \\
                    &     (0.138)         &     (0.145)         &     (0.146)         &     (0.410)         &     (0.072)         &     (0.089)         &     (0.082)         &     (0.153)         &     (0.136)         &     (0.168)         &     (0.141)         &     (0.241)         \\
Lagged Food Security (FS)&       0.110         &       0.023         &       0.218         &      -0.556         &       0.079         &      -0.047         &      -0.003         &      -0.101         &      -0.111         &       0.093         &      -0.133         &      -0.213         \\
                    &     (0.116)         &     (0.117)         &     (0.124)         &     (0.356)         &     (0.050)         &     (0.055)         &     (0.056)         &     (0.107)         &     (0.093)         &     (0.115)         &     (0.094)         &     (0.155)         \\
\midrule
Observations        &      12,031         &      12,031         &      12,031         &      12,031         &       9,303         &       9,303         &       9,303         &       9,303         &       3,277         &       3,277         &       3,277         &       3,277         \\
\hline \hline \\[-1.8ex] 
\multicolumn{13}{p{760pt}}{\small \noindent \textit{Note}: The table displays regression results 
from our dynamic panel specification with household fixed effects, round dummies, and region-time trends 
(see Equation \ref{eq:dyn}). FI stands for Fractional Index while FS stands for our standardized index of 
food insecurity. Standard errors, clustered at the household, are reported in parentheses 
(*** p$<$0.001, ** p$<$0.01, * p$<$0.05).}  \end{tabular}

%% file: tables/inc_bond_fs.tex
\begin{tabular}{l*{8}{c}} \\ [-1.8ex]\hline \hline \\[-1.8ex] 
& \multicolumn{4}{c}{Ethiopia} & \multicolumn{4}{c}{Malawi} \\ 
 & \multicolumn{1}{c}{FS Index} & \multicolumn{1}{c}{Mild} & 
\multicolumn{1}{c}{Moderate} & \multicolumn{1}{c}{Severe} & \multicolumn{1}{c}{FS Index} 
& \multicolumn{1}{c}{Mild} & \multicolumn{1}{c}{Moderate} & \multicolumn{1}{c}{Severe} 
\\ \midrule 
\midrule
Lagged Fractional Index (FI)&       0.548         &       0.568         &       0.591\sym{**} &       0.045         &      -0.051         &       1.161         &      -0.223         &       0.349         \\
                    &     (0.369)         &     (0.292)         &     (0.188)         &     (0.061)         &     (0.394)         &     (0.699)         &     (0.469)         &     (0.188)         \\
Lagged FS $\times$ Lagged FI&       0.262         &      -0.292         &      -0.218         &       0.203         &       0.276         &      -0.968         &       0.462         &      -0.311         \\
                    &     (0.230)         &     (0.308)         &     (0.256)         &     (0.381)         &     (0.215)         &     (0.696)         &     (0.476)         &     (0.288)         \\
Lagged Food Security (FS)&       0.407\sym{*}  &       0.666\sym{**} &       0.667\sym{**} &       0.143         &       0.401\sym{**} &       0.839         &       0.131         &       0.506\sym{**} \\
                    &     (0.201)         &     (0.249)         &     (0.214)         &     (0.317)         &     (0.130)         &     (0.430)         &     (0.294)         &     (0.185)         \\
\midrule
Observations        &       9,622         &       9,622         &       9,622         &       9,622         &       3,987         &       3,987         &       3,987         &       3,987         \\
\hline \hline \\[-1.8ex] 
\multicolumn{9}{p{560pt}}{\small \noindent \textit{Note}: The table displays regression results 
from our dynamic panel specification with household fixed effects, round dummies, and region-time trends. 
FI stands for Fractional Index while Increased (Decreased) indicates how the household's income changed 
in the past 30 days. Standard errors, clustered at the household, are reported in parentheses 
(*** p$<$0.001, ** p$<$0.01, * p$<$0.05).}  \end{tabular}

%% file: tables/inter_fs.tex
\begin{tabular}{l*{12}{c}} \\ [-1.8ex]\hline \hline \\[-1.8ex] 
& \multicolumn{4}{c}{Ethiopia} & \multicolumn{4}{c}{Malawi} & 
\multicolumn{4}{c}{Nigeria} \\ & \multicolumn{1}{c}{FS Index} & \multicolumn{1}{c}{Mild} & 
\multicolumn{1}{c}{Moderate} & \multicolumn{1}{c}{Severe} & \multicolumn{1}{c}{FS Index} 
& \multicolumn{1}{c}{Mild} & \multicolumn{1}{c}{Moderate} & \multicolumn{1}{c}{Severe} 
& \multicolumn{1}{c}{FS Index} & \multicolumn{1}{c}{Mild} & \multicolumn{1}{c}{Moderate} 
& \multicolumn{1}{c}{Severe} \\ \midrule 
\midrule
Lagged Fractional Index (FI)&       7.570\sym{*}  &       2.792         &       0.723         &       1.288         &       0.312         &      -0.795         &       0.505         &      -0.185         &       1.100         &       0.423         &       0.131         &       0.271         \\
                    &     (3.036)         &     (2.887)         &     (1.754)         &     (0.832)         &     (0.430)         &     (0.577)         &     (0.421)         &     (0.219)         &     (0.644)         &     (0.670)         &     (0.480)         &     (0.225)         \\
Lagged FS $\times$ Lagged FI&       5.086         &      -2.887         &       7.670         &      14.879         &      -0.466         &       0.927         &      -0.486         &      -0.442         &      -0.502         &      -0.348         &       0.358         &      -0.360         \\
                    &     (3.054)         &     (3.623)         &     (4.371)         &     (8.628)         &     (0.393)         &     (0.615)         &     (0.539)         &     (0.919)         &     (0.540)         &     (0.710)         &     (0.607)         &     (1.079)         \\
Lagged Food Security (FS)&      -3.837         &       2.239         &      -6.233         &     -11.453         &       0.579\sym{*}  &      -0.041         &       0.593         &       0.515         &       0.383         &       0.539         &      -0.032         &      -0.193         \\
                    &     (2.578)         &     (2.820)         &     (3.626)         &     (7.484)         &     (0.273)         &     (0.420)         &     (0.371)         &     (0.682)         &     (0.354)         &     (0.455)         &     (0.384)         &     (0.712)         \\
COVID-19 Stringency &       0.120\sym{***}&       0.069\sym{*}  &       0.013         &       0.012         &       0.015\sym{*}  &       0.000         &       0.008         &       0.004         &       0.021\sym{*}  &       0.011         &       0.007         &       0.003         \\
                    &     (0.034)         &     (0.030)         &     (0.019)         &     (0.009)         &     (0.007)         &     (0.007)         &     (0.005)         &     (0.003)         &     (0.010)         &     (0.006)         &     (0.006)         &     (0.002)         \\
Lagged FS $\times$ COVID-19 Stringency&       0.050         &      -0.028         &       0.081         &       0.138         &      -0.009         &       0.001         &      -0.011         &      -0.011         &      -0.007         &      -0.006         &      -0.002         &       0.000         \\
                    &     (0.033)         &     (0.036)         &     (0.046)         &     (0.095)         &     (0.005)         &     (0.007)         &     (0.006)         &     (0.012)         &     (0.006)         &     (0.006)         &     (0.006)         &     (0.013)         \\
Lagged FI $\times$ COVID-19 Stringency&      -0.097\sym{*}  &      -0.035         &      -0.009         &      -0.018         &      -0.003         &       0.015         &      -0.008         &       0.003         &      -0.011         &      -0.006         &      -0.002         &      -0.003         \\
                    &     (0.038)         &     (0.036)         &     (0.022)         &     (0.011)         &     (0.008)         &     (0.010)         &     (0.007)         &     (0.004)         &     (0.009)         &     (0.008)         &     (0.006)         &     (0.003)         \\
Lagged FS $\times$ COVID-19 Stringency $\times$ Lagged FI&      -0.066         &       0.035         &      -0.101         &      -0.182         &       0.008         &      -0.017         &       0.009         &       0.009         &       0.010         &       0.005         &      -0.002         &       0.007         \\
                    &     (0.039)         &     (0.046)         &     (0.055)         &     (0.110)         &     (0.007)         &     (0.011)         &     (0.009)         &     (0.016)         &     (0.008)         &     (0.009)         &     (0.009)         &     (0.019)         \\
\midrule
Observations        &      12,031         &      12,031         &      12,031         &      12,031         &       9,303         &       9,303         &       9,303         &       9,303         &       3,277         &       3,277         &       3,277         &       3,277         \\
\hline \hline \\[-1.8ex] 
\multicolumn{13}{p{850pt}}{\small \noindent \textit{Note}: The table displays regression results 
from our dynamic panel specification with household fixed effects, round dummies, and region-time trends 
(see Equation \ref{eq:dynint}). FI stands for Fractional Index while FS stands for our standardized index of 
food insecurity. Standard errors, clustered at the household, are reported in parentheses 
(*** p$<$0.001, ** p$<$0.01, * p$<$0.05).}  \end{tabular}

%% file: tables/anc_fs.tex
\begin{tabular}{l*{11}{c}} \\ [-1.8ex]\hline \hline \\[-1.8ex] 
& \multicolumn{4}{c}{Ethiopia} & \multicolumn{3}{c}{Malawi} & 
\multicolumn{4}{c}{Nigeria} \\ & \multicolumn{1}{c}{FS Index} & \multicolumn{1}{c}{Mild} & 
\multicolumn{1}{c}{Moderate} & \multicolumn{1}{c}{Severe} & \multicolumn{1}{c}{FS Index} 
& \multicolumn{1}{c}{Mild} & \multicolumn{1}{c}{Moderate} 
& \multicolumn{1}{c}{FS Index} & \multicolumn{1}{c}{Mild} & \multicolumn{1}{c}{Moderate} 
& \multicolumn{1}{c}{Mild} \\ \midrule 
\midrule
Baseline HHI        &      -0.107         &       0.014         &      -0.039         &      -0.031         &      -0.101         &      -0.175\sym{*}  &      -0.182\sym{**} &       0.017         &      -0.019         &      -0.013         &      -0.013         \\
                    &     (0.106)         &     (0.042)         &     (0.043)         &     (0.022)         &     (0.116)         &     (0.072)         &     (0.055)         &     (0.107)         &     (0.062)         &     (0.061)         &     (0.048)         \\
Baseline FS $\times$ Baseline HHI&      -0.066         &       0.017         &       0.084         &      -0.204\sym{***}&       0.412\sym{*}  &       0.087         &       0.270\sym{*}  &      -0.012         &       0.005         &       0.051         &      -0.259         \\
                    &     (0.126)         &     (0.067)         &     (0.119)         &     (0.042)         &     (0.182)         &     (0.082)         &     (0.113)         &     (0.095)         &     (0.074)         &     (0.084)         &     (0.220)         \\
Baseline Food Security (FS)&       0.287\sym{***}&       0.152\sym{**} &       0.103         &       0.148\sym{***}&       0.176         &       0.075         &       0.018         &       0.189\sym{*}  &       0.072         &       0.108         &       0.311         \\
                    &     (0.085)         &     (0.051)         &     (0.084)         &     (0.031)         &     (0.117)         &     (0.053)         &     (0.076)         &     (0.074)         &     (0.058)         &     (0.066)         &     (0.164)         \\
\midrule
Observations        &      14,506         &      14,506         &      14,506         &      14,506         &      14,370         &      14,370         &      14,370         &       5,335         &       5,335         &       5,335         &       5,335         \\
\hline \hline \\[-1.8ex] 
\multicolumn{12}{p{700pt}}{\small \noindent \textit{Note}: The table displays regression results 
from our ANCOVA specification with round and regioncontrols 
(see Equation \ref{eq:anc}). HHI stands for Herfindahl-Hirschman Index  while FS stands for our standardized index of 
food insecurity. Note that in the Malawi baseline data no respondent reported being severely food insecure and so we cannot estimate the ANCOVA specification for severe food insecurity in Malawi. Standard errors, clustered at the household, are reported in parentheses 
(*** p$<$0.001, ** p$<$0.01, * p$<$0.05).}  \end{tabular}

%% file: tables/indices_sum_app.tex
\begin{longtable}{ P{1cm} P{2cm} P{1.5cm} P{8.5cm} P{8.5cm}} \caption{Livelihood Diversification Indices Summary}   \label{tab:indices_app} \\ [-1.8ex]\hline \hline

Index Type &  Standardized Across Countries & Time Period & Description & Pre-COVID-19 Kernel Density Graph \\ \hline
\endfirsthead 
\hline \hline Index Type &  Standardized Across Countries & Time Period & Description & Pre-COVID-19 Kernel Density Graph \\ \hline
\endhead
\multicolumn{5}{c}{{Continued on Next Page\ldots}} 
\endfoot 
\endlastfoot 
\centering
Fraction & No &	Pre- and Post-COVID-19 & For this index we collapse variables into income categories that are consistent across rounds within a country but vary across countries. As a result, this index allow us to observe income sources at the most granular level available over multiple waves for each country individually. This fraction index considers 10 income source categories in Ethiopia, 7 in Malawi and Nigeria, and 8 in Uganda. & 
    \includegraphics[width=85mm, scale=1]{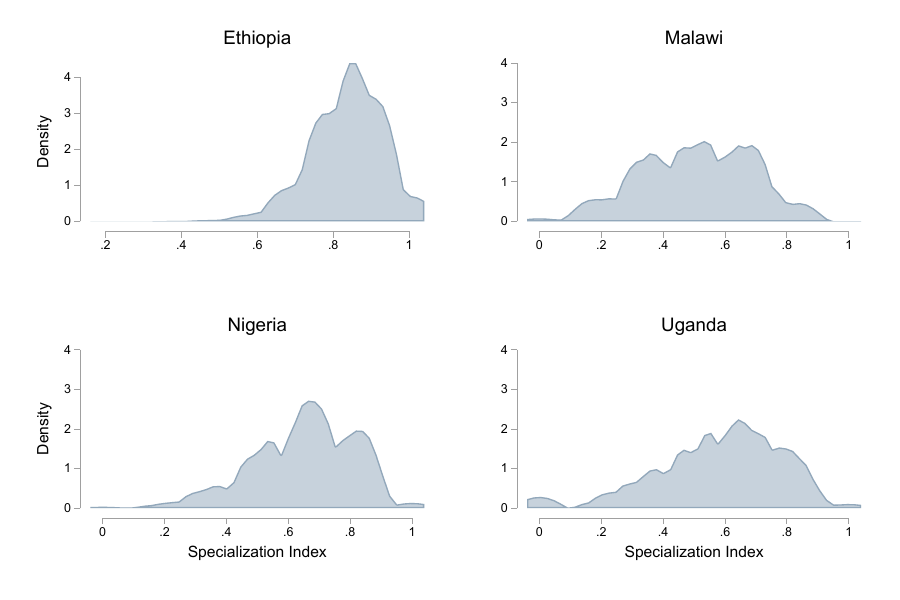} \\
\midrule
Fraction &	Yes &	Pre-COVID-19 &	For this index we collapse pre-COVID variables into income categories that are consistent across countries. Because the index only draws from the rich pre-COVID data, we are able to include 12 income categories available across all four countries: remittances; in-kind assistance from friends and family; investments and savings; income from properties; pension; non-farm enterprise; crop sales and consumption; livestock sales; livestock products sales and consumption; wages; government and NGO assistance; and other. &
    \includegraphics[width=85mm, scale=1]{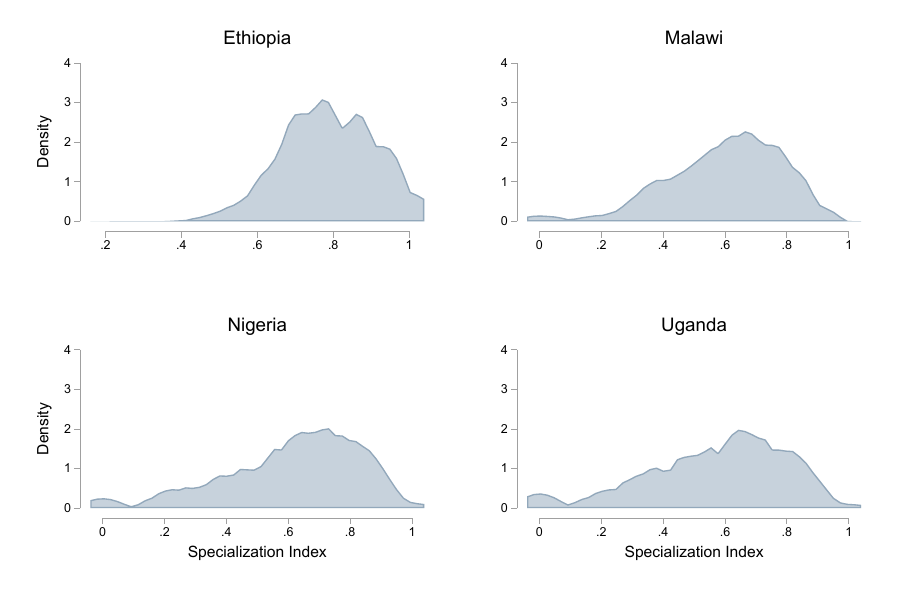} \\
\midrule 
Fraction &	No &	Pre-COVID-19 &	For this index we collapse pre-COVID variables into income categories that vary across country. As a result, this index allow us to observe income sources at the most granular level available in the pre-COVID data. We generate this index using 19 income sources in Ethiopia, 21 in Malawi, 15 in Nigeria, and 13 in Uganda.  & 
    \includegraphics[width=85mm, scale=1]{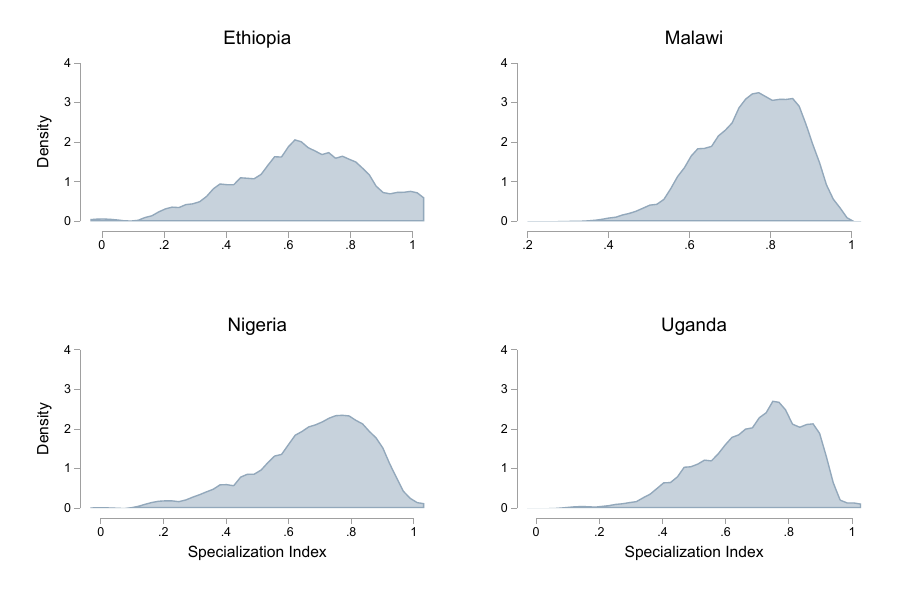} \\
    \midrule
HHI &	No &	Pre-COVID-19 &	This index is identical to the fractional Pre-COVID-19 index described above but uses an HHI instead of a fraction to evaluate the distribution of income from each source. The index varies across country. & 
    \includegraphics[width=85mm, scale=1]{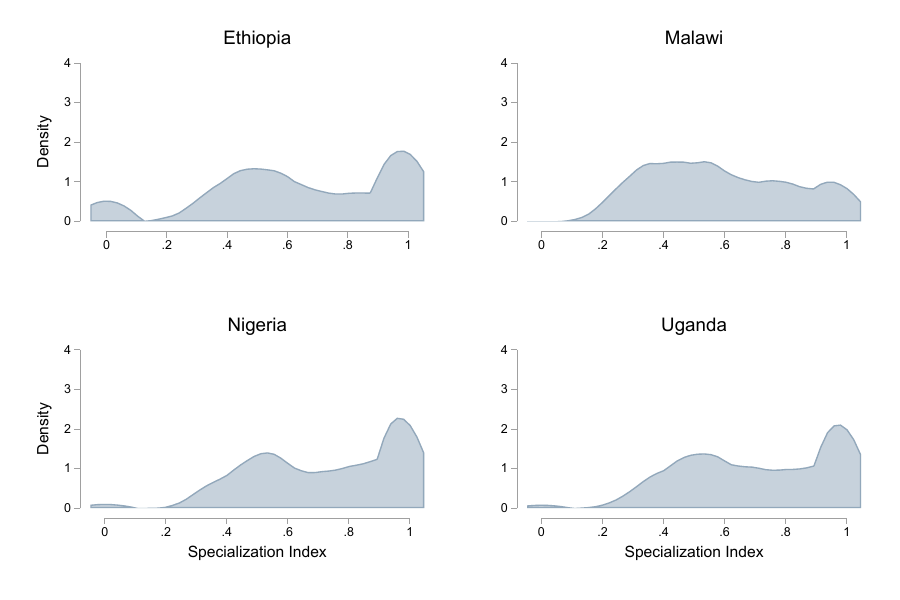} \\
\midrule \midrule 
\multicolumn{5}{p{23cm}}{\footnotesize \textit{Note}: The table summarizes the four livelihood diversification indices that we pre-specified but did not include in our main analysis. Higher index values indicate more household specialization (less income diversification).}
\end{longtable}

%% file: tables/app_index2.tex
\begin{tabular}{P{10cm} P{10cm}}  \\ [-1.8ex]\hline \hline \\[-1.8ex] 
\includegraphics[width=100mm, scale=1]{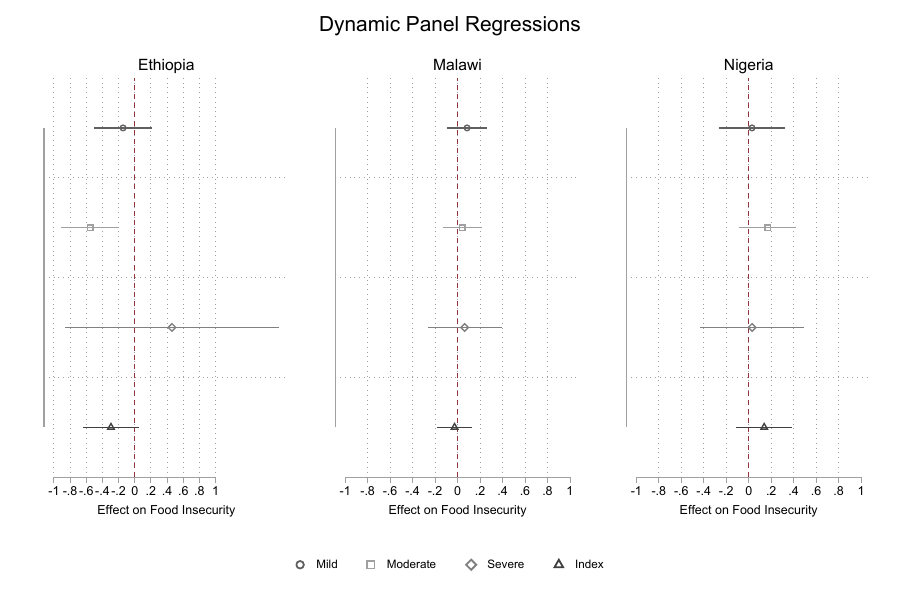} & \includegraphics[width=100mm, scale=1]{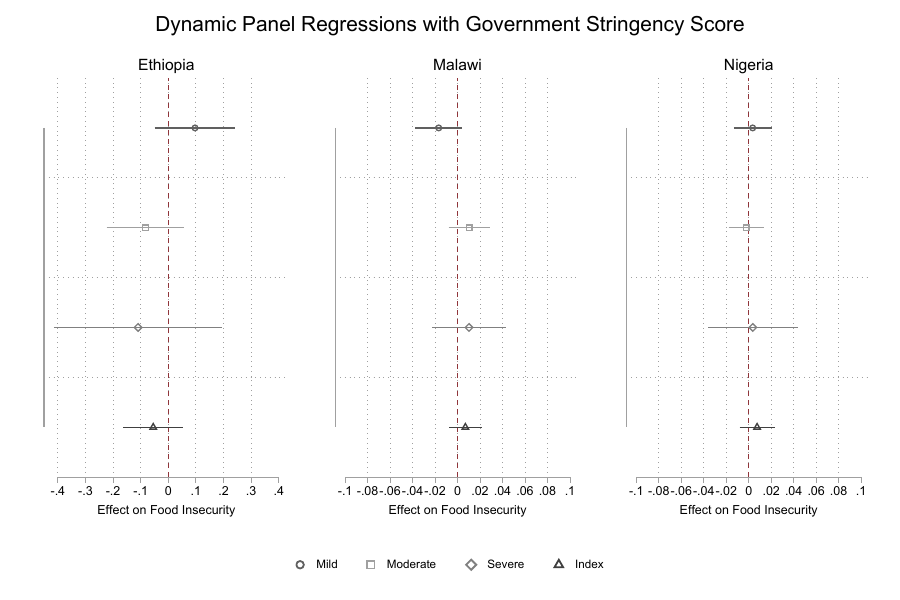} \\
\midrule
\midrule
\multicolumn{2}{J{20cm}}{\noindent \textit{Note:} The figures plot results using our second fractional index that spans pre- and post-outbreak data. This index differs from our main fractional index in that it is not standardized across countries. Results come from dynamic panel and dynamic panel with interactions regression (see Equation \ref{eq:anc}). Horizontal lines represent 95 percent confidence intervals calculated using standard errors clustered at the household.}
\end{tabular}

%% file: tables/app_anc_fs.tex
\begin{tabular}{P{10cm} P{10cm}}  \\ [-1.8ex]\hline \hline \\[-1.8ex] 
\includegraphics[width=100mm, scale=1]{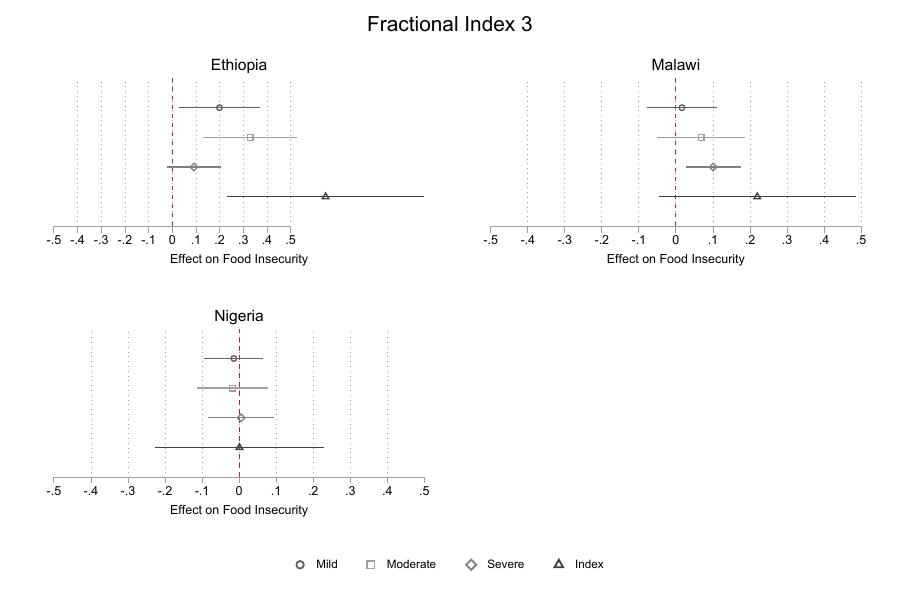} & \includegraphics[width=100mm, scale=1]{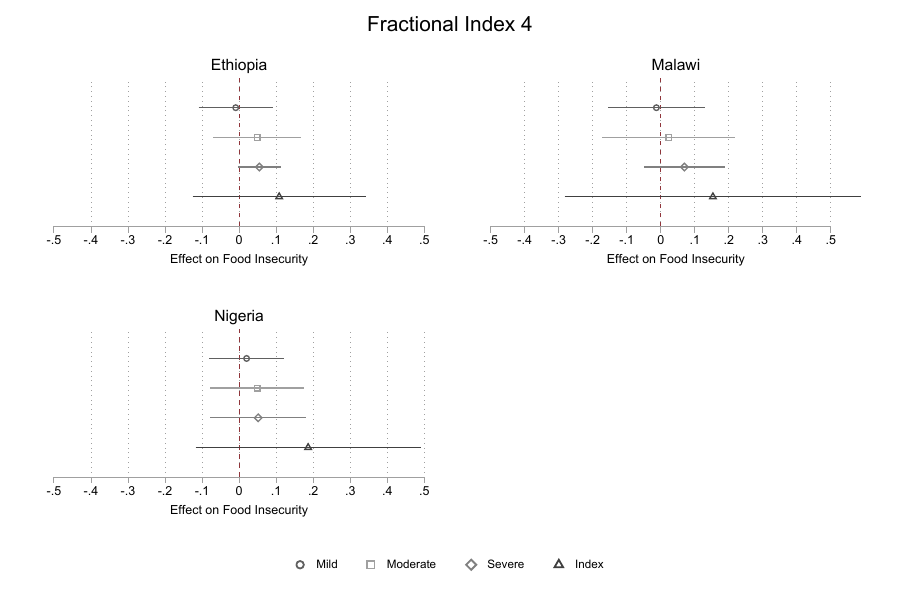} \\
\includegraphics[width=100mm, scale=1]{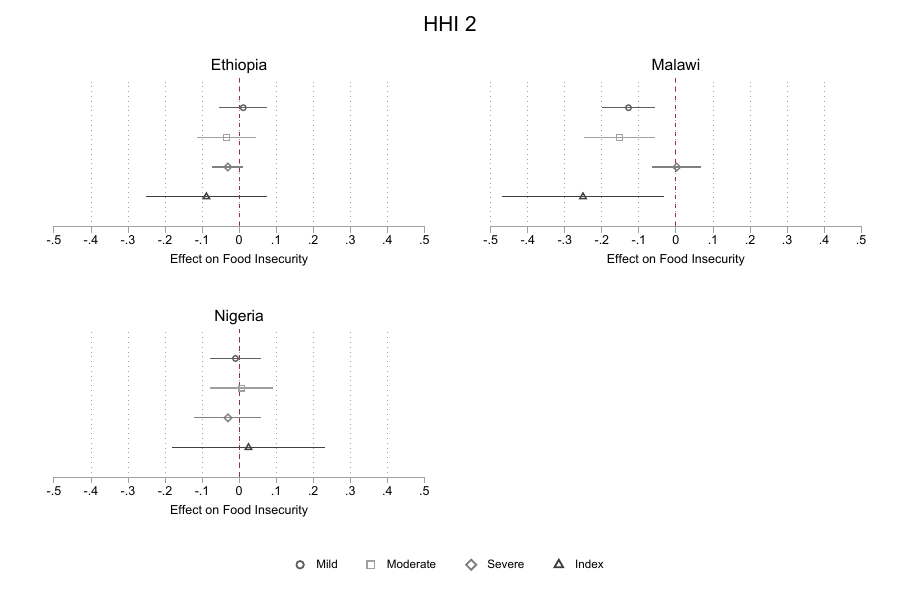} &  \\
\midrule
\midrule
\multicolumn{2}{J{20cm}}{\noindent \textit{Note:} The figures plot results using our second fractional index that spans pre- and post-outbreak data. This index differs from our main fractional index in that it is not standardized across countries. Results come from dynamic panel and dynamic panel with interactions regression (see Equation \ref{eq:anc}). Horizontal lines represent 95 percent confidence intervals calculated using standard errors clustered at the household.}
\end{tabular}

%% file: tables/app_did_fs.tex
\begin{tabular}{P{10cm} P{10cm}}  \\ [-1.8ex]\hline \hline \\[-1.8ex] 
\includegraphics[width=100mm, scale=1]{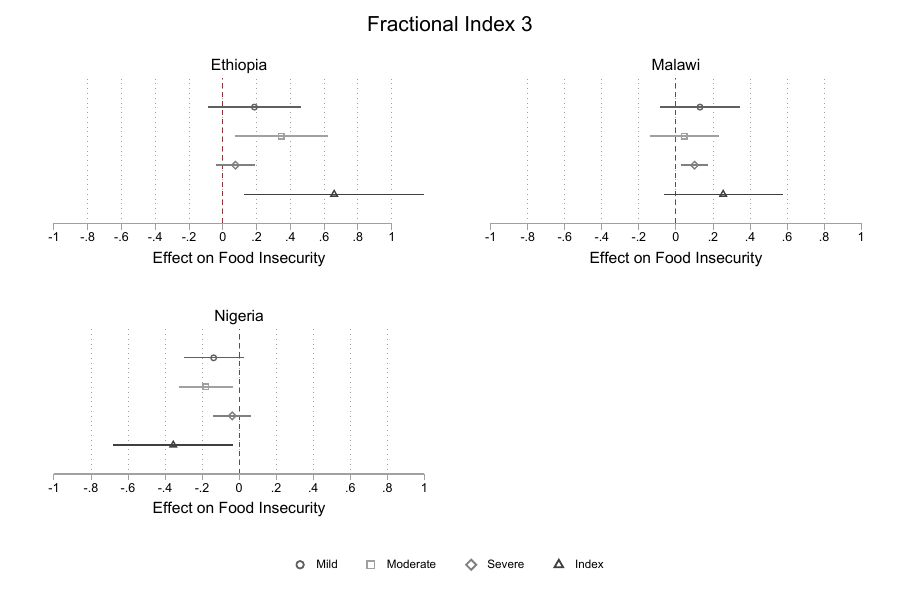} & \includegraphics[width=100mm, scale=1]{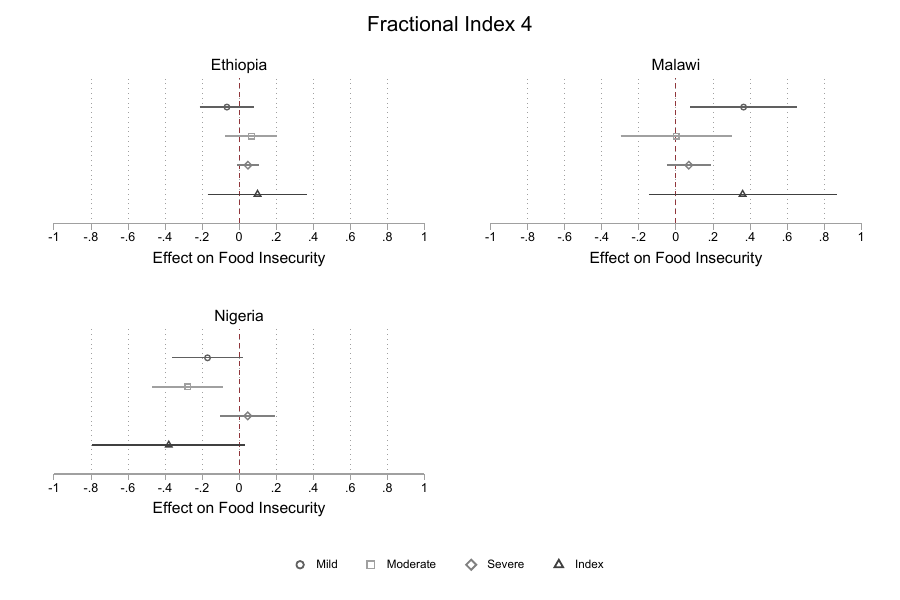} \\
\includegraphics[width=100mm, scale=1]{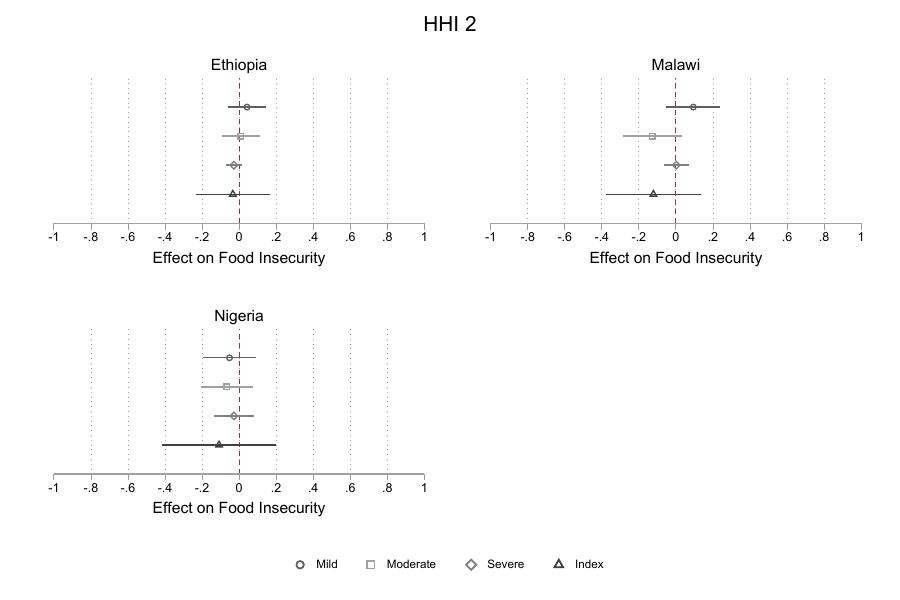} &  \\
\midrule
\midrule
\multicolumn{2}{J{20cm}}{\noindent \textit{Note:} The figures plot results using our second fractional index that spans pre- and post-outbreak data. This index differs from our main fractional index in that it is not standardized across countries. Results come from dynamic panel and dynamic panel with interactions regression (see Equation \ref{eq:anc}). Horizontal lines represent 95 percent confidence intervals calculated using standard errors clustered at the household.}
\end{tabular}

%% file: tables/ind_density_sec_sex.tex
\begin{longtable}{P{8cm} P{8cm}} \caption{Pre-COVID-19 Indices Density by Urban/Rural and Head-of-Household Gender}   \label{tab:ind_sec_sex} \\ [-1.8ex]\hline \hline 

Urban/Rural  & Male-Headed/Female-Headed \\ \hline
\endfirsthead 

\hline \hline 
Urban/Rural  & Male-Headed/Female-Headed \\ \hline
\endhead

\multicolumn{2}{c}{{Continued on Next Page\ldots}} 
\endfoot 
\endlastfoot 
\centering
\includegraphics[width=77mm, scale=1]{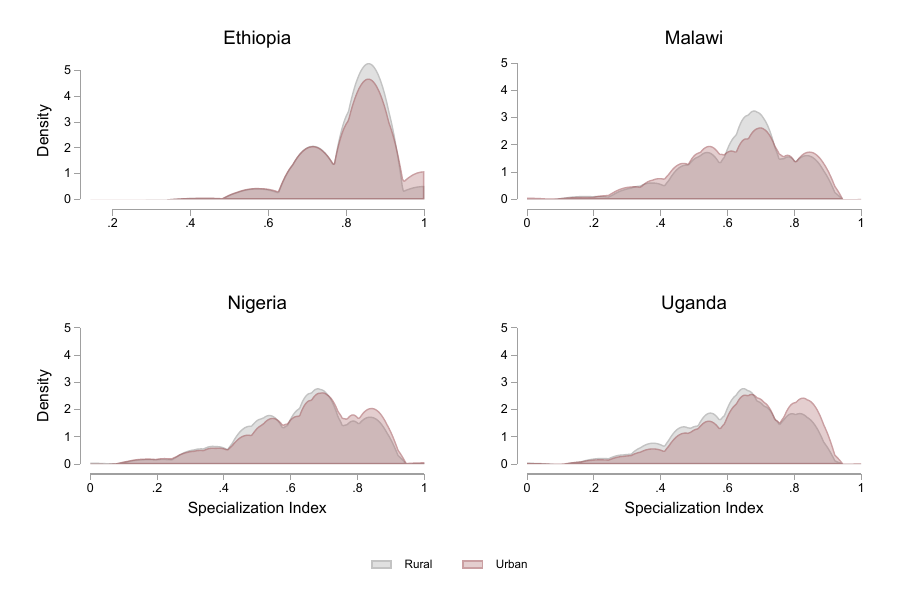} & \includegraphics[width=77mm, scale=1]{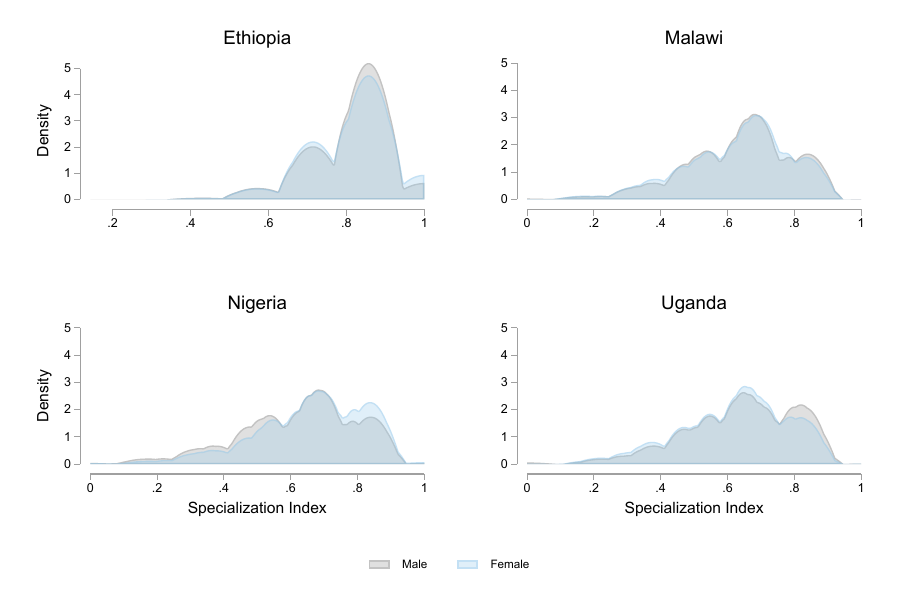} \\
\multicolumn{2}{c}{Fractional Index 1} \\
\midrule
\includegraphics[width=77mm, scale=1]{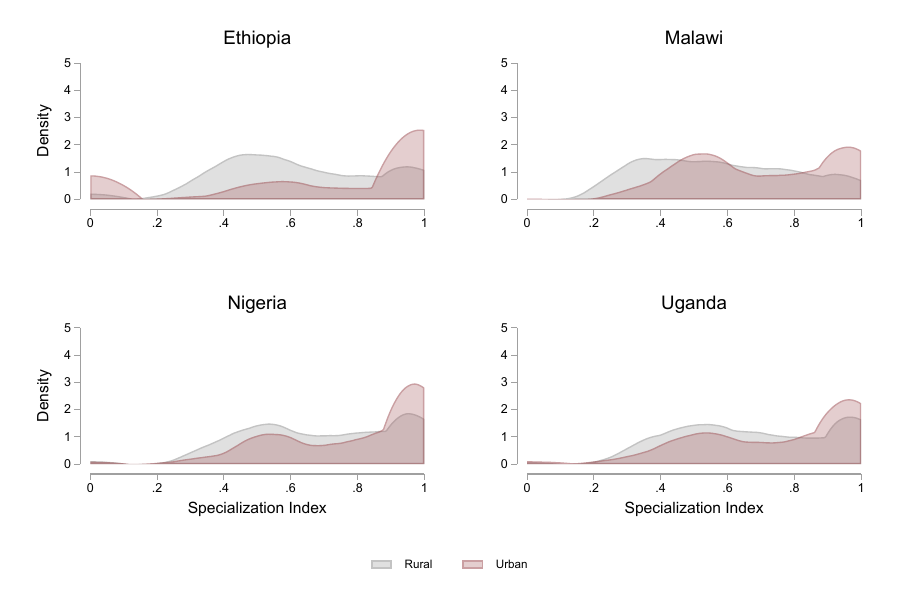} & \includegraphics[width=77mm, scale=1]{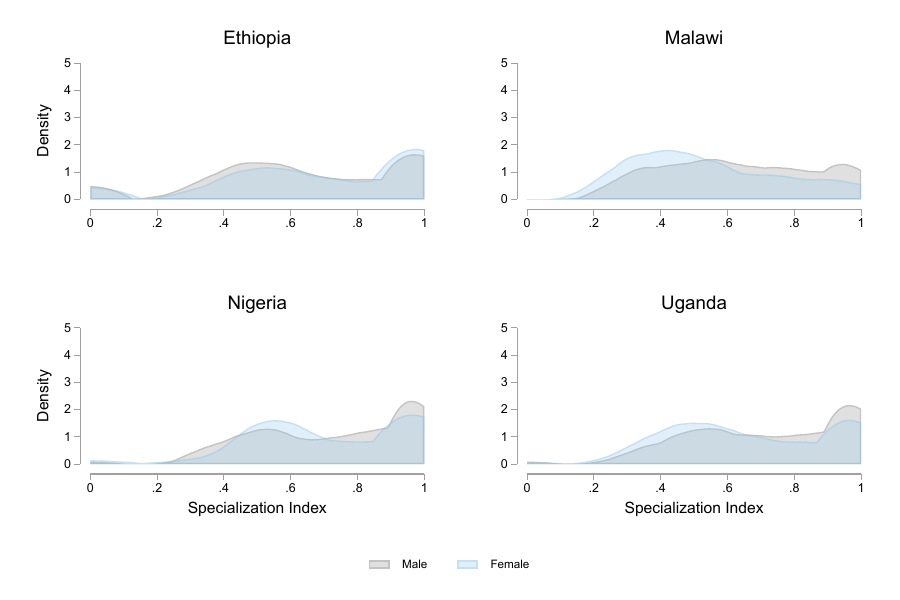} \\
\multicolumn{2}{c}{HHI 1} \\
\midrule
\includegraphics[width=77mm, scale=1]{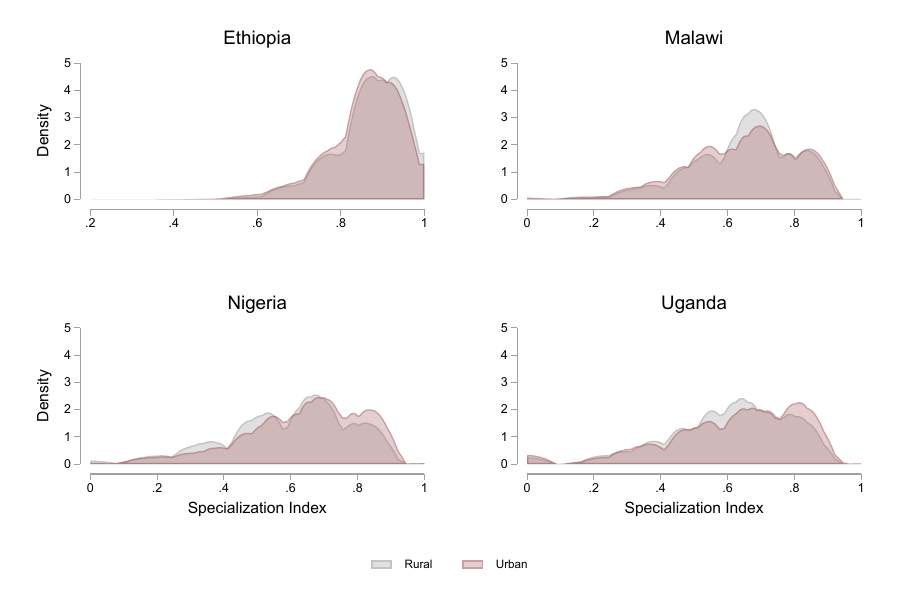} & \includegraphics[width=77mm, scale=1]{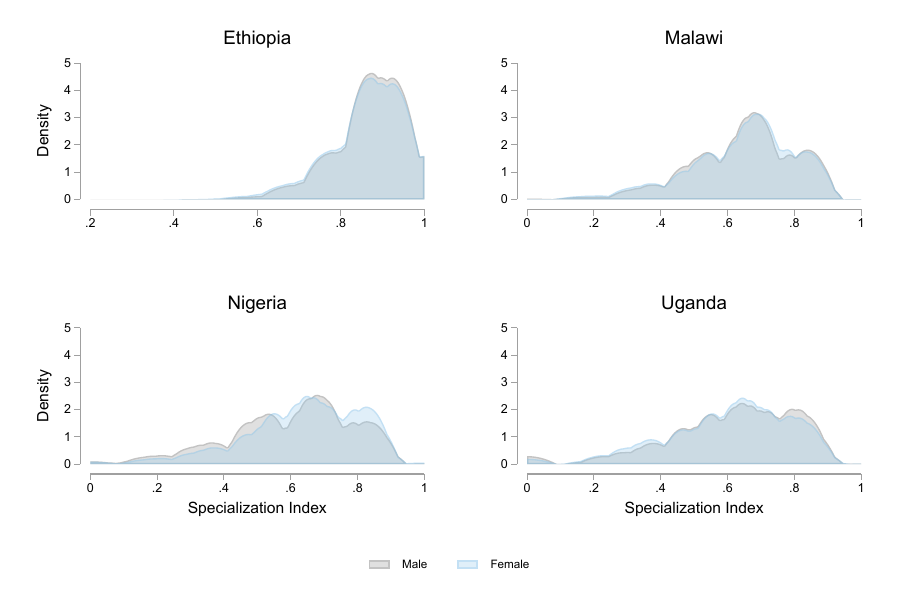} \\
\multicolumn{2}{c}{Fractional Index 2} \\
\midrule
\includegraphics[width=77mm, scale=1]{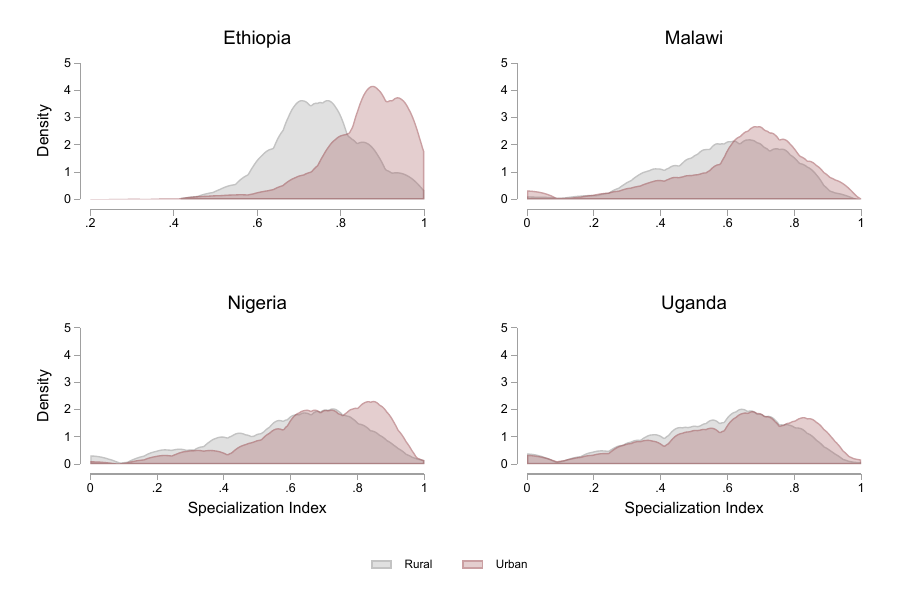} & \includegraphics[width=77mm, scale=1]{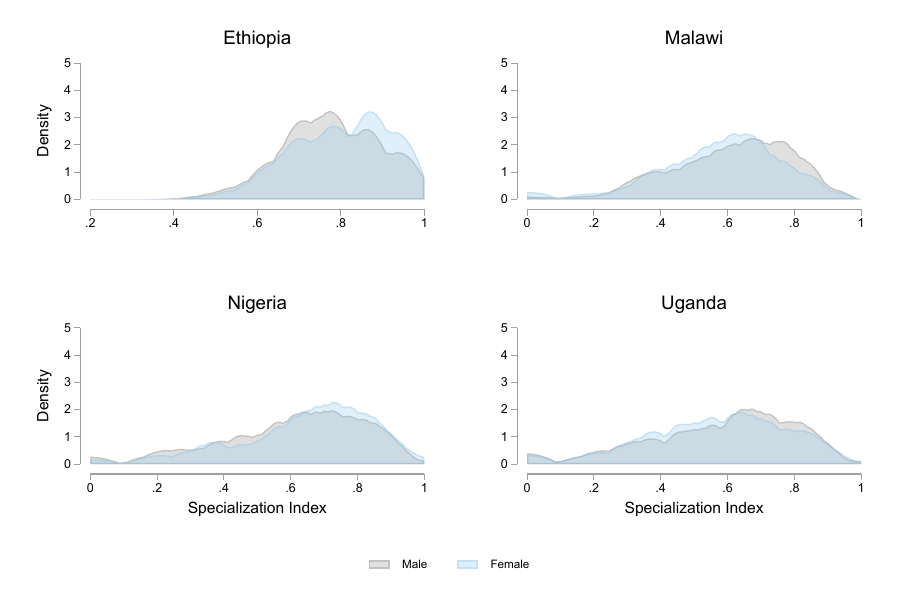} \\
\multicolumn{2}{c}{Fractional Index 3} \\
\midrule
\includegraphics[width=77mm, scale=1]{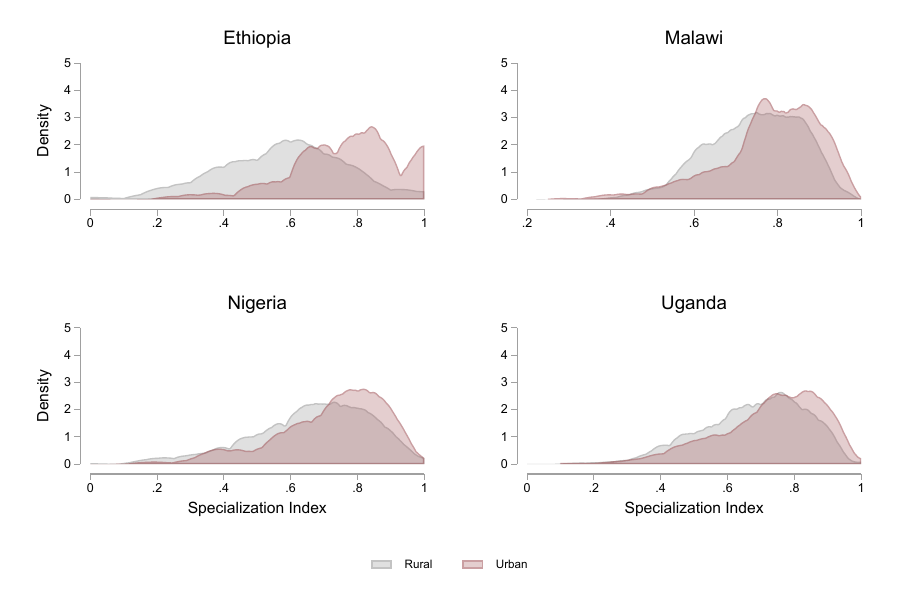} & \includegraphics[width=77mm, scale=1]{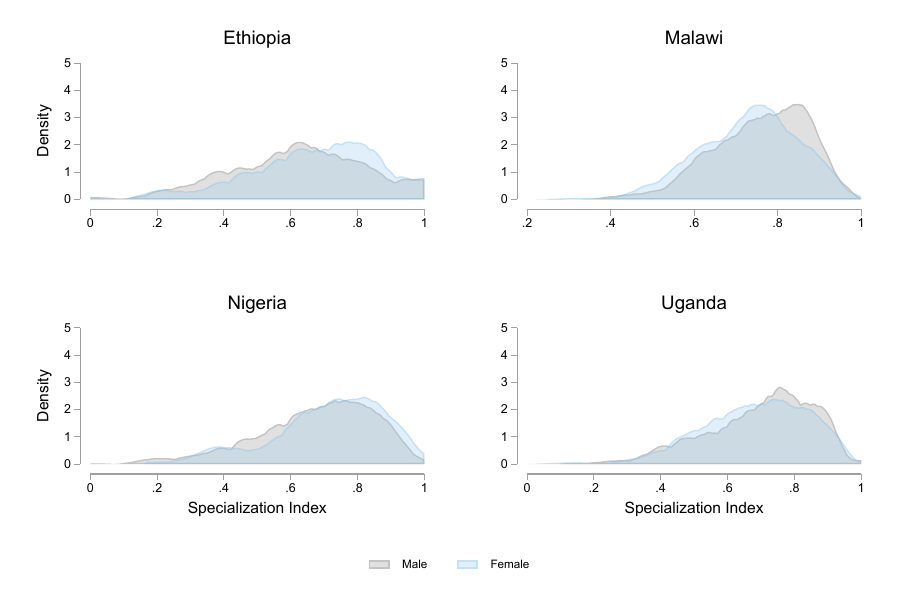} \\
\multicolumn{2}{c}{Fractional Index 4} \\
\midrule
\includegraphics[width=77mm, scale=1]{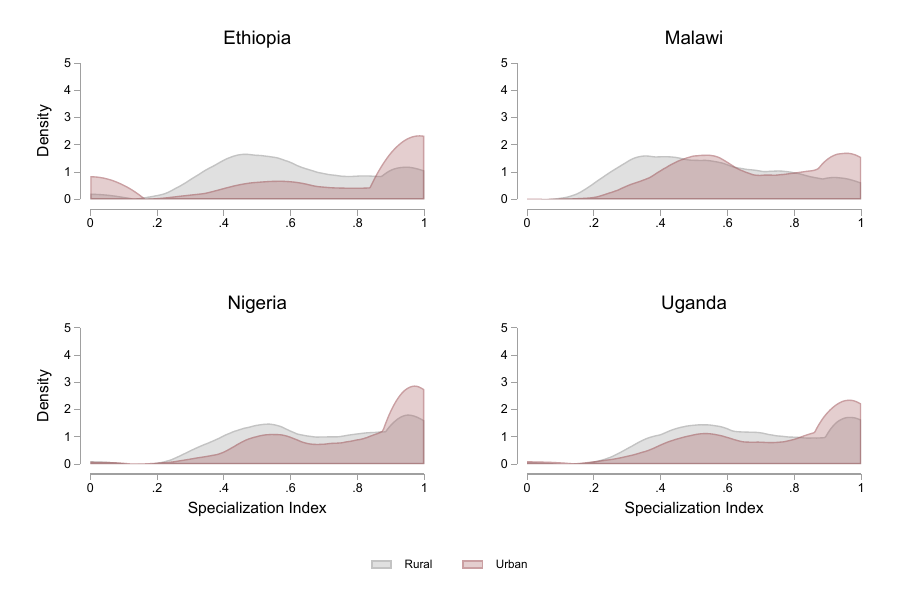} & \includegraphics[width=77mm, scale=1]{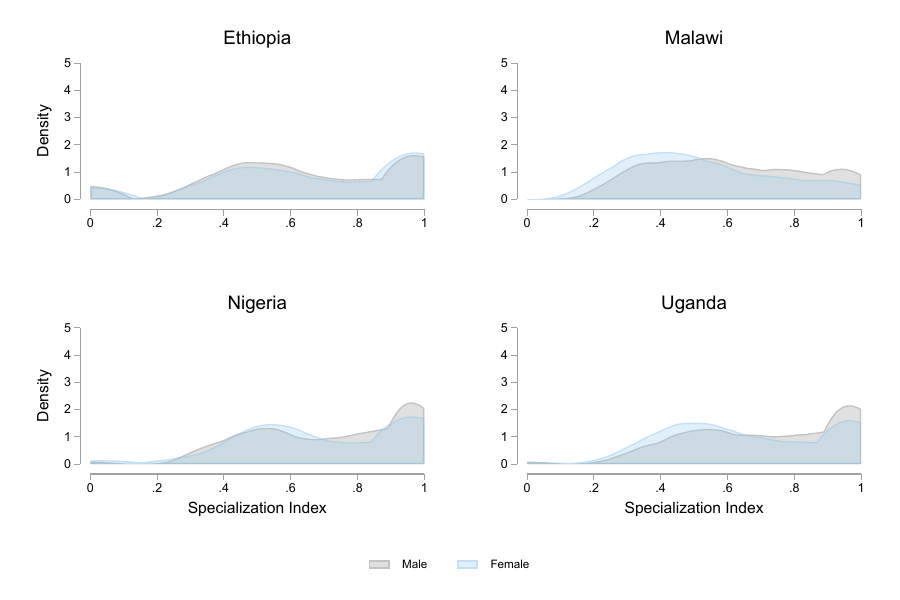} \\
\multicolumn{2}{c}{HHI 2} \\
\hline \hline
\multicolumn{2}{J{17cm}}{\textit{Note:} The table displays kernel density graphs for the pre-COVID-19 round for each of the six livelihood diversification indices. The first column of graphs shows densities for urban versus rural populations while the second column of graphs separates the data by male- versus female-headed households. Higher index values indicate more household specialization (less income diversification).}
\end{longtable}

%% file: tables/fs_anc_sex.tex
\begin{tabular}{P{10cm} P{10cm}}  \\ [-1.8ex]\hline \hline \\[-1.8ex] 
\centering
\includegraphics[width=100mm, scale=1]{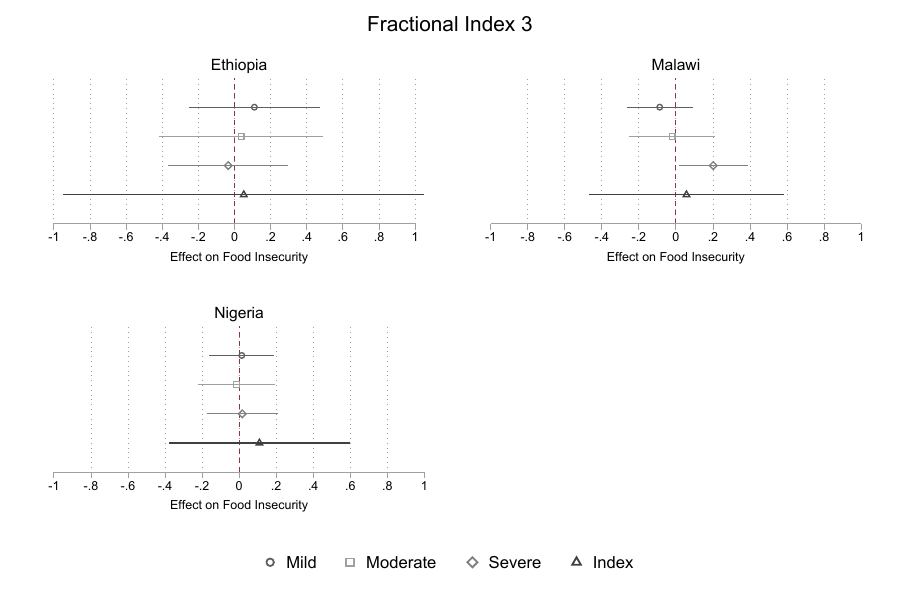} & \includegraphics[width=100mm, scale=1]{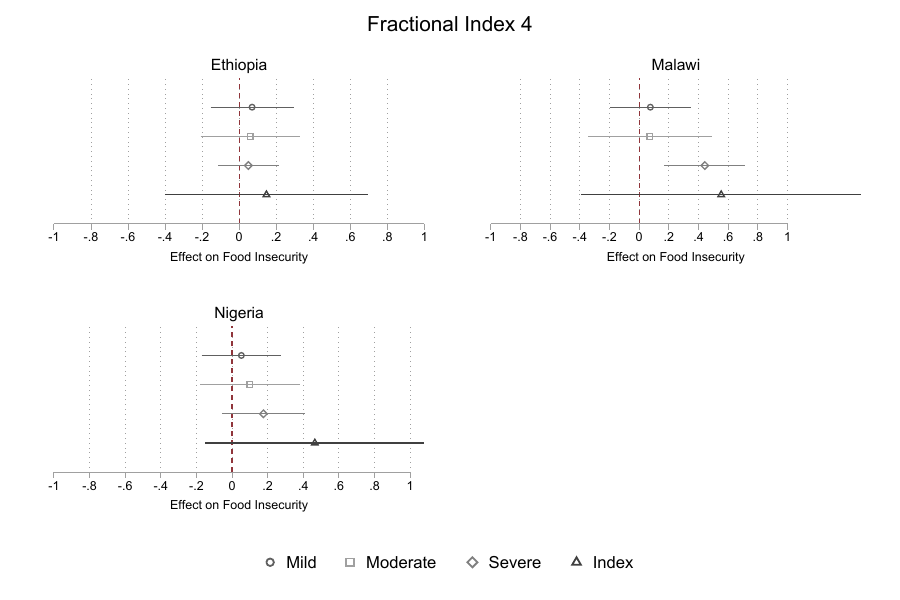} \\
\midrule
\includegraphics[width=100mm, scale=1]{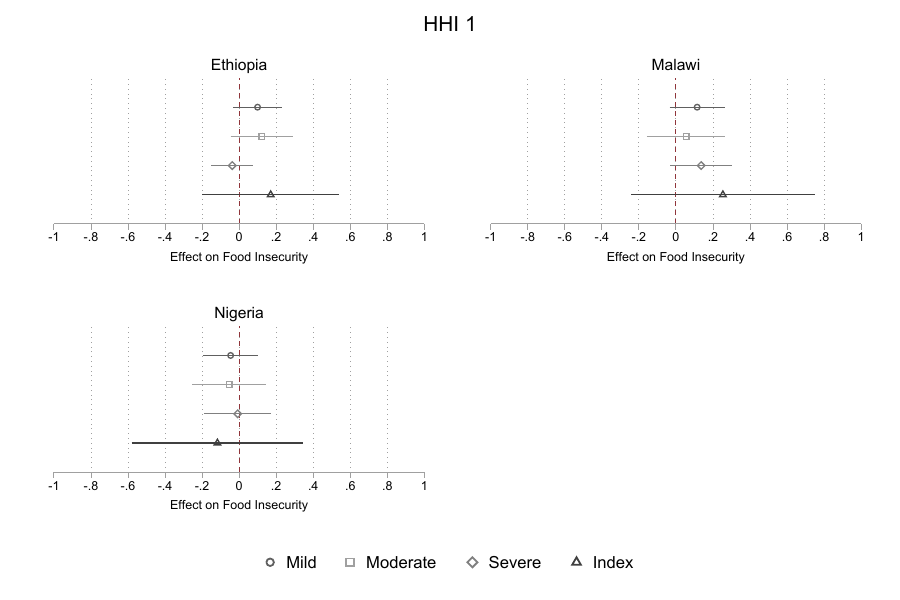} & \includegraphics[width=100mm, scale=1]{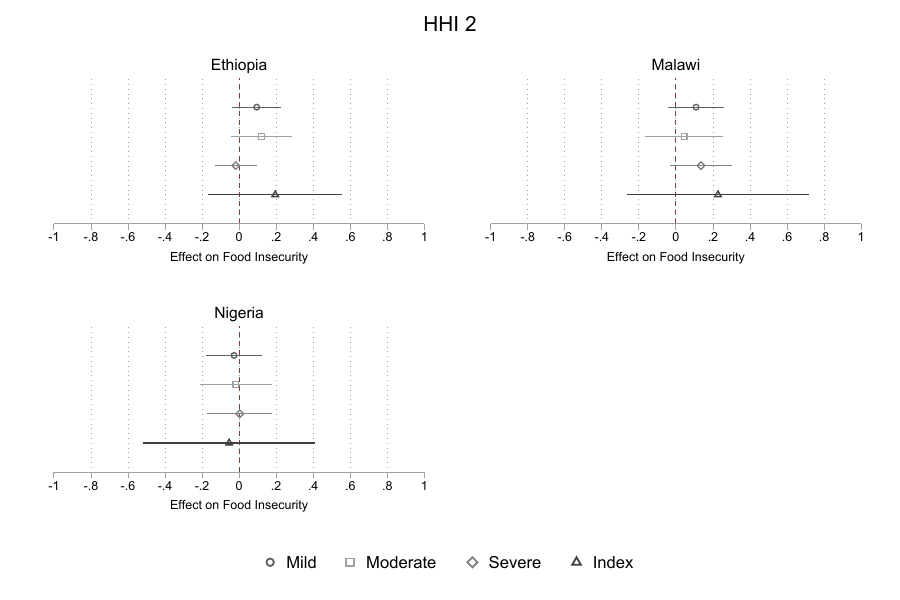} \\
\midrule
\midrule
\multicolumn{2}{J{20cm}}{\noindent \textit{Note:} The figure plots ANCOVA regression results with region and round controls and standard errors clustered at the household level (see Equation \ref{eq:anc_app}). We display coefficients for the interaction of lagged income diversity indices (Indices 3-6) and a head-of-household gender indicator. Male-headed households serve as the comparison group. Horizontal lines represent 95 percent confidence intervals.}
\end{tabular}

%% file: tables/fs_did_sex.tex
\begin{tabular}{P{10cm} P{10cm}}  \\ [-1.8ex]\hline \hline \\[-1.8ex] 
\centering
\includegraphics[width=100mm, scale=1]{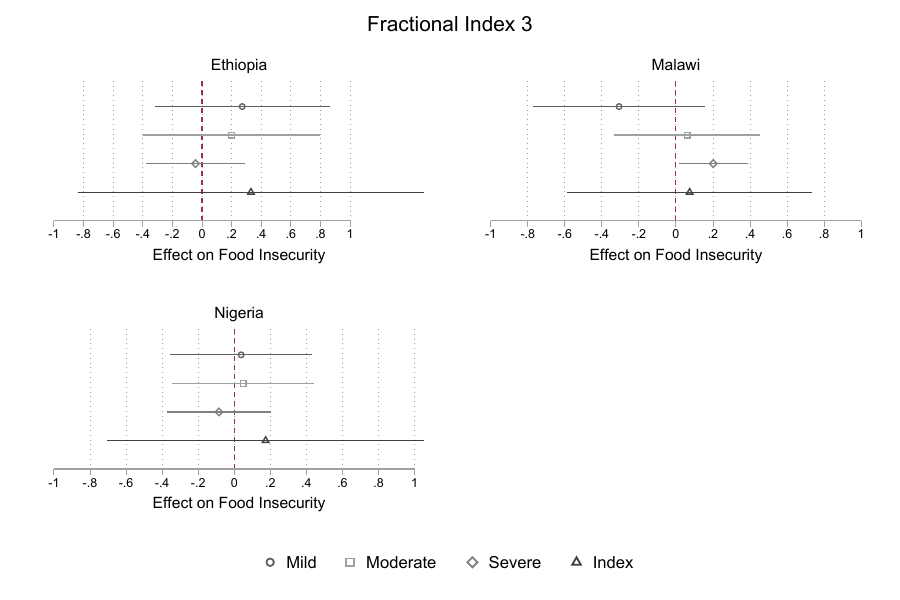} & \includegraphics[width=100mm, scale=1]{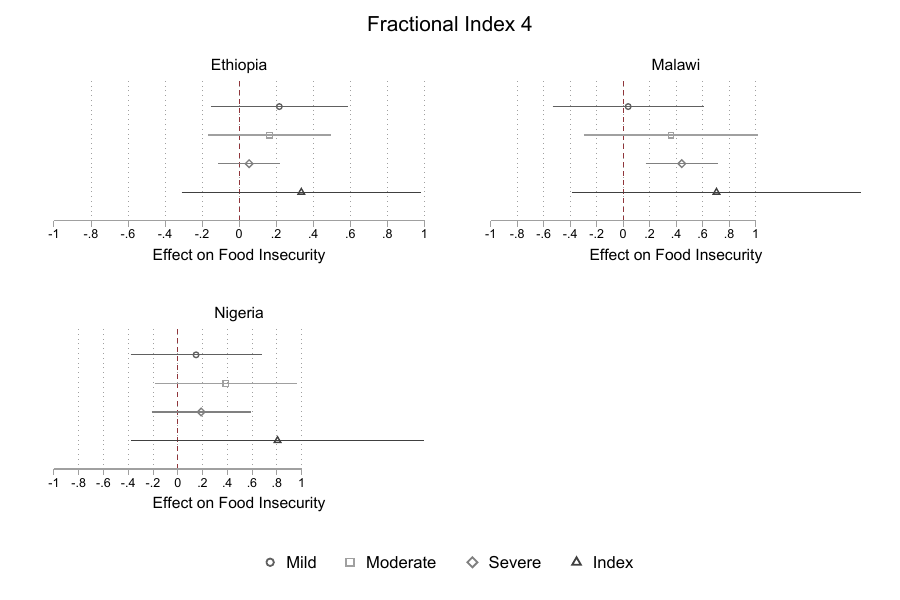} \\
\midrule
\includegraphics[width=100mm, scale=1]{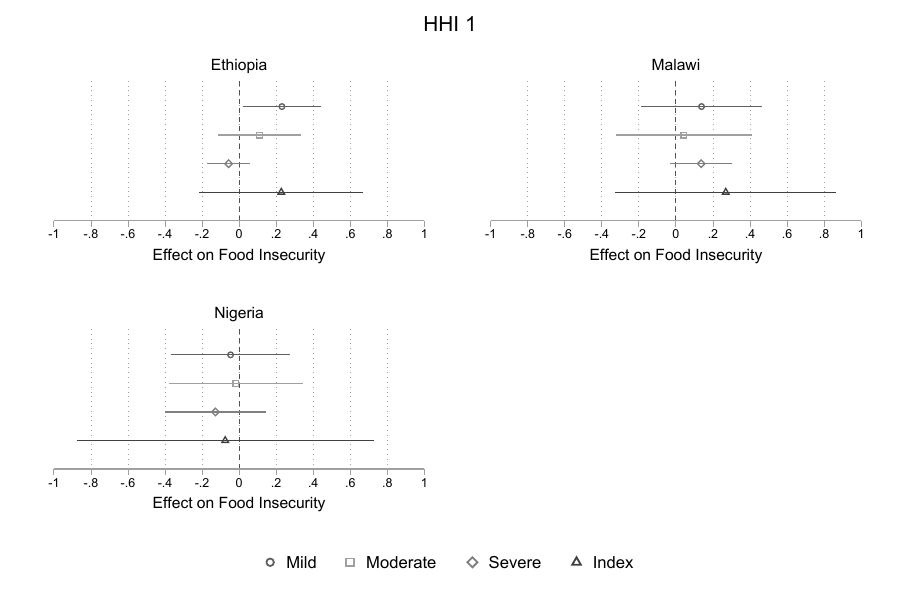} & \includegraphics[width=100mm, scale=1]{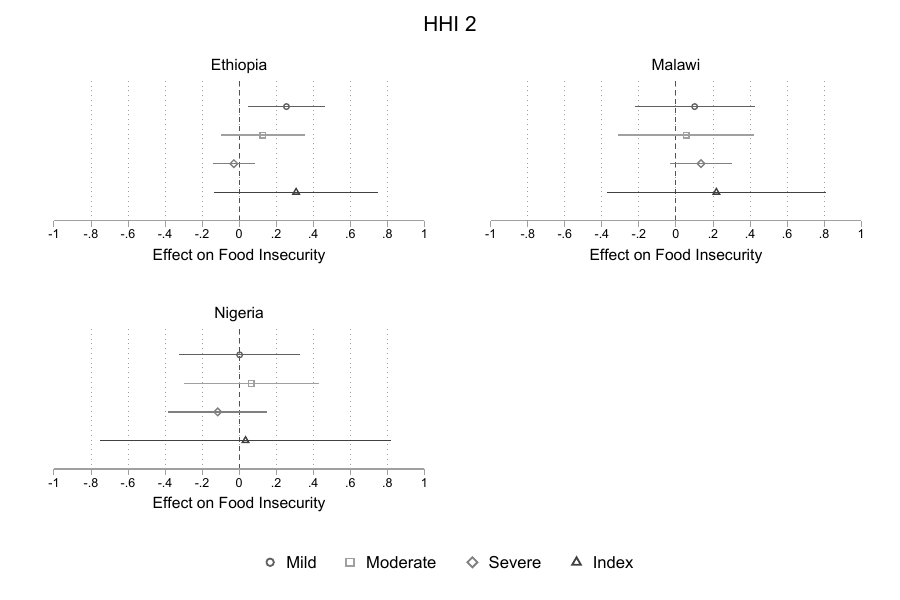} \\
\midrule
\midrule
\multicolumn{2}{J{20cm}}{\noindent \textit{Note:} The figure plots difference-in-difference regression results with region and round controls and standard errors clustered at the household level (see Equation \ref{eq:did_app}). We display coefficients for the interaction of lagged income diversity indices (Indices 3-6) and a head-of-household gender indicator. Male-headed households serve as the comparison group. Horizontal lines represent 95 percent confidence intervals.}
\end{tabular}

%% file: tables/fs_anc_sec.tex
\begin{tabular}{P{10cm} P{10cm}}  \\ [-1.8ex]\hline \hline \\[-1.8ex] 
\centering
\includegraphics[width=100mm, scale=1]{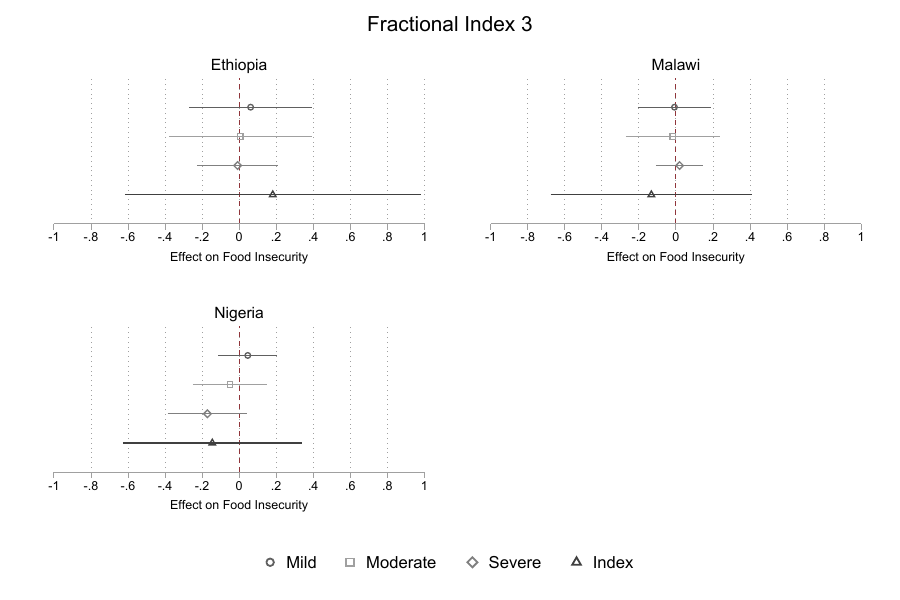} & \includegraphics[width=100mm, scale=1]{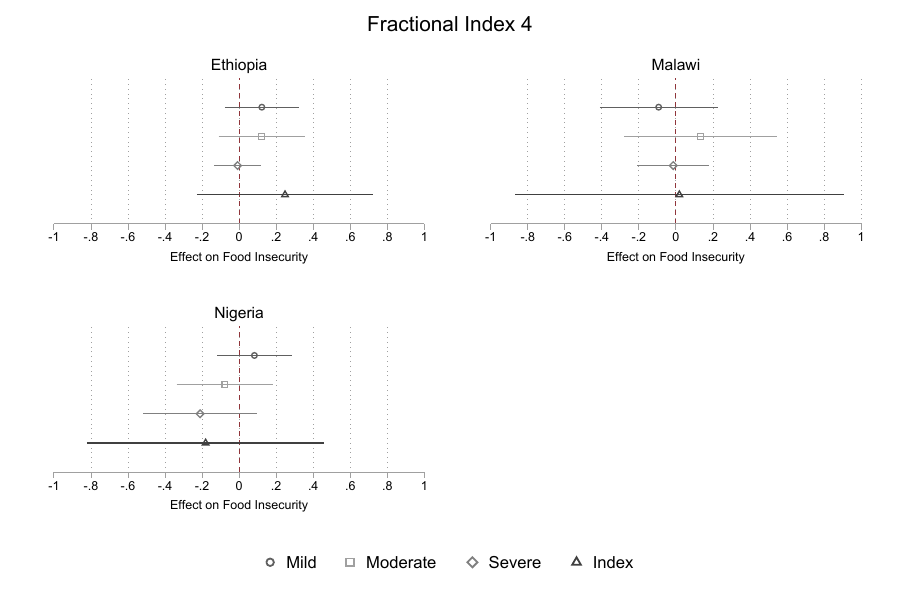} \\
\midrule
\includegraphics[width=100mm, scale=1]{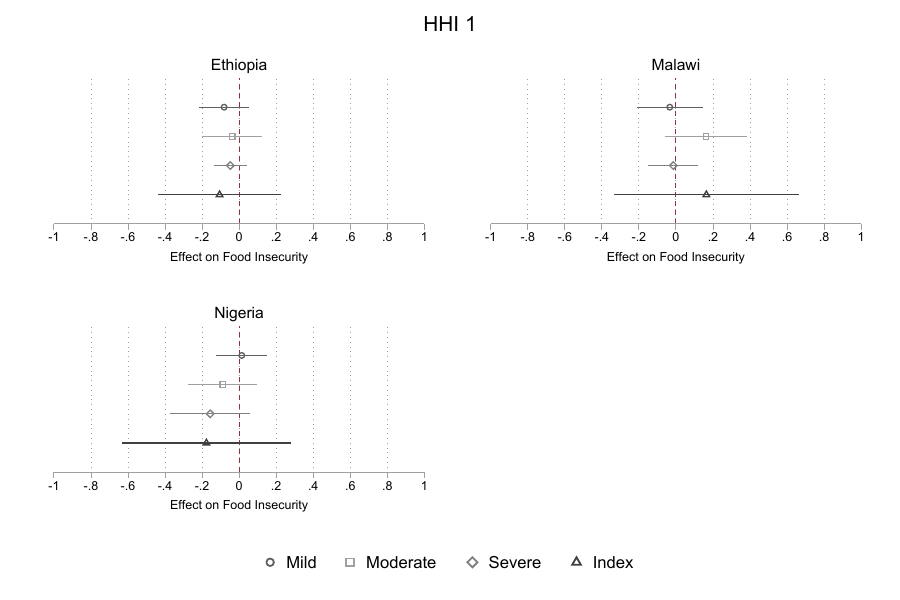} & \includegraphics[width=100mm, scale=1]{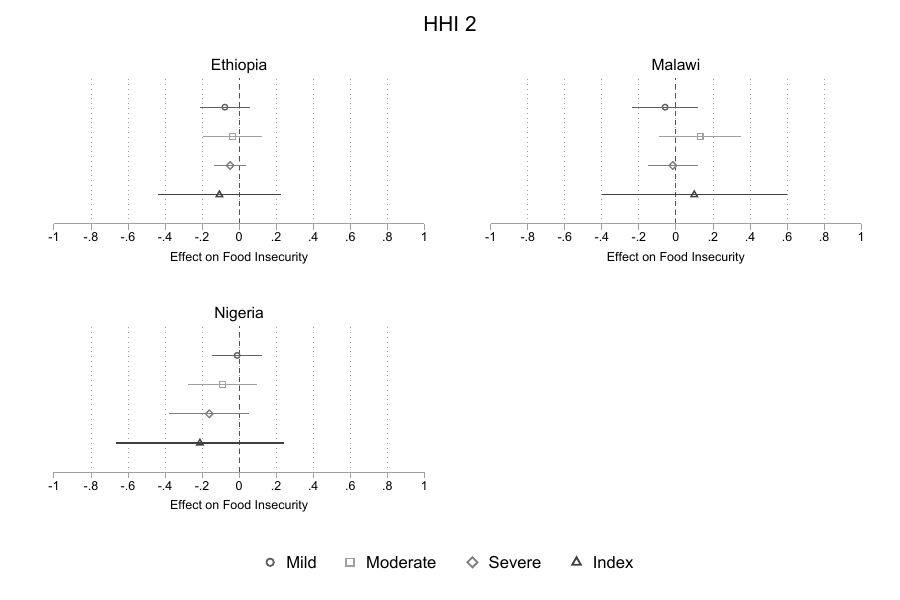} \\
\midrule
\midrule
\multicolumn{2}{J{20cm}}{\noindent \textit{Note:} The figure plots ANCOVA regression results with region and round controls and standard errors clustered at the household level (see Equation \ref{eq:anc_app}). We display coefficients for the interaction of lagged income diversity indices (Indices 3-6) and a head-of-household gender indicator. Male-headed households serve as the comparison group. Horizontal lines represent 95 percent confidence intervals.}
\end{tabular}

%% file: tables/fs_did_sec.tex
\begin{tabular}{P{10cm} P{10cm}}  \\ [-1.8ex]\hline \hline \\[-1.8ex] 
\centering
\includegraphics[width=100mm, scale=1]{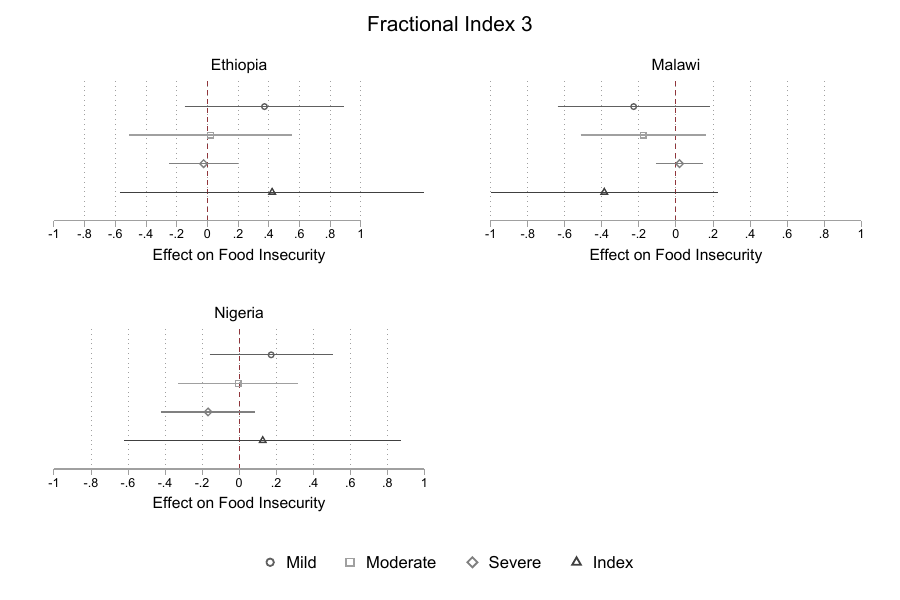} & \includegraphics[width=100mm, scale=1]{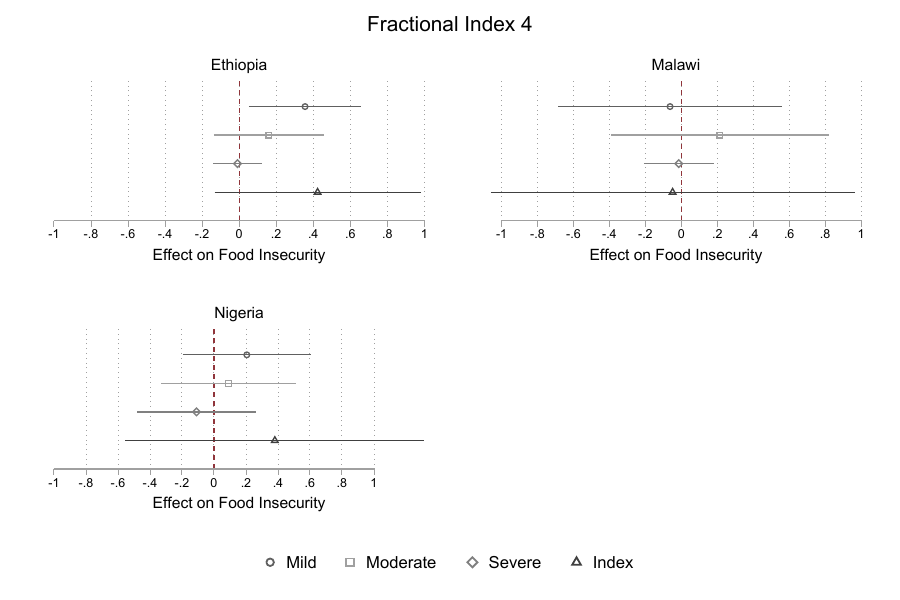} \\
\midrule
\includegraphics[width=100mm, scale=1]{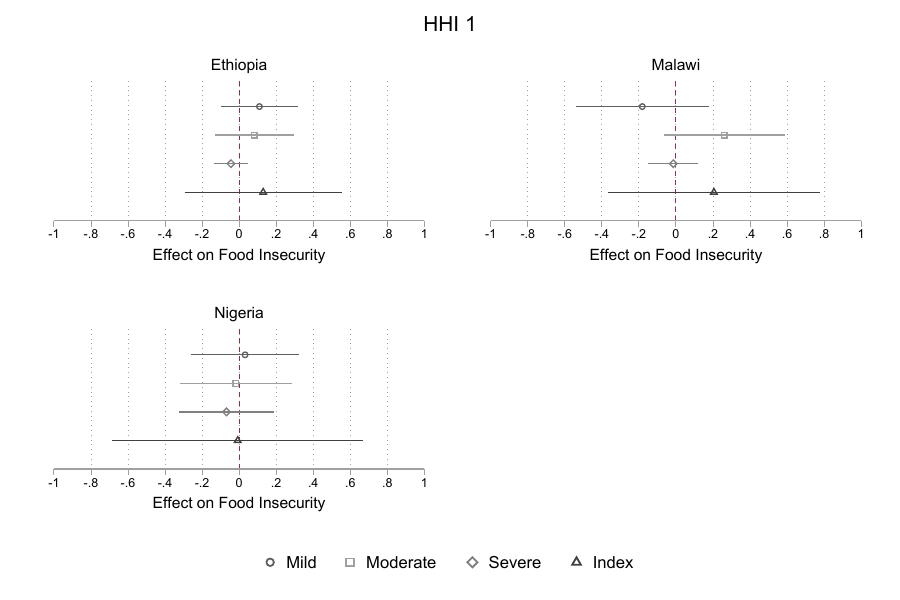} & \includegraphics[width=100mm, scale=1]{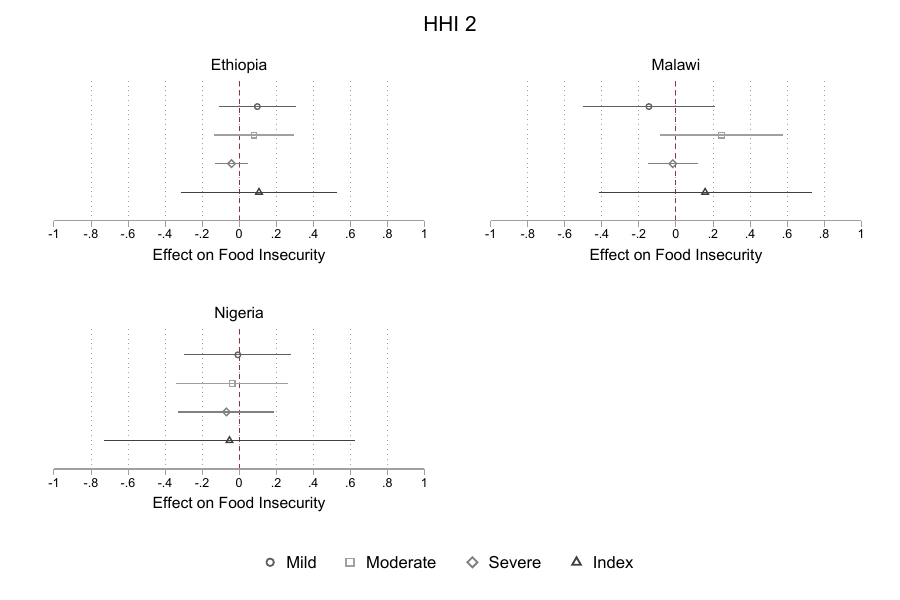} \\
\midrule
\midrule
\multicolumn{2}{J{20cm}}{\noindent \textit{Note:} The figure plots difference-in-difference regression results with region and round controls and standard errors clustered at the household level (see Equation \ref{eq:did_app}). We display coefficients for the interaction of lagged income diversity indices (Indices 3-6) and a head-of-household gender indicator. Male-headed households serve as the comparison group. Horizontal lines represent 95 percent confidence intervals.}
\end{tabular}